\documentclass[aps,twocolumn,preprintnumbers,amsmath,amssymb,superscriptaddress]{revtex4-1}

\usepackage{amsmath,amssymb,amsfonts,graphicx,multirow,color,bbm,textcomp,mathtools,dsfont,bm}
\usepackage{paralist}

\usepackage{srcltx}
\usepackage[all,cmtip]{xy}
\usepackage{mathrsfs}
\usepackage{srcltx,soul,subfigure}
\usepackage{txfonts}

\newcommand{\ket}[1]{| #1 \rangle}

\newcommand{\bee}{\begin{equation}}
\newcommand{\ee}{\end{equation}}
\newcommand{\bma}{\begin{pmatrix}}
\newcommand{\ema}{\end{pmatrix}}
\newcommand{\balig}{\begin{align}}
\newcommand{\ealig}{\end{align}}

\newcommand{\ba}{\begin{align}}
\newcommand{\ea}{\end{align}}

\newcommand{\ignore}[1]{}

\newcommand{\clr}{\color{red}}

\newcommand{\br}{{\bm{r}}}

\newcommand{\bk}{{\bm{k}}}
\newcommand{\bmu}{\bm{u}}
\newcommand{\bnu}{\bm{\nu}}

\usepackage{siunitx}

\begin{document}


\title{Scalable Majorana vortex modes in iron-based superconductors}

\author{Ching-Kai Chiu}\email{Corresponding: qiujingkai@ucas.edu.cn}
\affiliation{Kavli Institute for Theoretical Sciences, University of Chinese Academy of Sciences, Beijing 100190, China}

\author{T. Machida}
\affiliation{RIKEN Center for Emergent Matter Science, Wako, Saitama 351-0198, Japan}

\author{Yingyi Huang}
\affiliation{Kavli Institute for Theoretical Sciences, University of Chinese Academy of Sciences, Beijing 100190, China}
\affiliation{Institute for Advanced Study, Tsinghua University, Beijing, 100084, China}

\author{T. Hanaguri}
\affiliation{RIKEN Center for Emergent Matter Science, Wako, Saitama 351-0198, Japan}

\author{Fu-Chun Zhang}
\affiliation{Kavli Institute for Theoretical Sciences, University of Chinese Academy of Sciences, Beijing 100190, China}
\affiliation{CAS Center for Excellence in Topological Quantum Computation, University of Chinese Academy of Sciences, Beijing 100190, China}
\affiliation{Collaborative Innovation Center of Advanced Microstructures, Nanjing University, Nanjing 210093, China}

\date{\today}

\begin{abstract} 
The iron-based superconductor FeTe$_x$Se$_{1-x}$ is one of the material candidates hosting Majorana vortex modes residing in the vortex cores. It has been observed by recent scanning tunneling spectroscopy measurement that the fraction of vortex cores possessing zero-bias peaks decreases with increasing magnetic field on the surface of FeTe$_x$Se$_{1-x}$. The hybridization of two Majorana vortex modes cannot simply explain this phenomenon. We construct a three-dimensional tight-binding model simulating the physics of over a hundred Majorana vortex modes in FeTe$_x$Se$_{1-x}$. Our simulation shows that the Majorana hybridization and disordered vortex distribution can explain the decreasing fraction of the zero-bias peaks observed in the experiment; the statistics of the energy peaks off zero energy in our Majorana simulation are in agreement with the experiment. These agreements lead to an important indication of scalable Majorana vortex modes in FeTe$_x$Se$_{1-x}$. Thus, FeTe$_x$Se$_{1-x}$ can be one promising platform possessing scalable Majorana qubits for quantum computing.  
\end{abstract}

\maketitle








\noindent
\textbf{\large Introduction}

\noindent
Majorana zero modes, being their own antiparticle in many-body systems, have attracted great attention for quantum computing~\cite{Kitaev2001,RMP_braiding}. Two of the promising platforms to host these exotic modes are a one-dimensional superconducting proximitized semiconductor nanowire~\cite{Gil_Majorana_wire,Sau_semiconductor_heterostructures,Roman_SC_semi} and an s-wave superconducting surface with a Dirac cone~\cite{Fu2008Superconducting}. The observation of the quantized Majorana conductance~\cite{Zhang:2018aa}  in the nanowire and the distinct zero bias peaks (ZBP)~\cite{Wang2018Evidence,2018arXiv181208995M,PhysRevX.8.041056} in the vortex cores brings excitement and encouragement in this field. The Majorana zero modes may form qubits storing quantum information; the key property of the distant Majorana modes for quantum computing is the quantum information in the Majorana qubits can be protected by the topology. On the other hand, according to the DiVincenzo criteria~\cite{DiVincenzo:2000aa}, achieving quantum computing requires a platform hosting scalable qubits. Seeking for scalable qubit systems is one of the major obstacles to pursue quantum computing. 


  

It has been observed that the iron-based superconductors (${\rm FeTe}_{x}{\rm Se}_{1-x}$, $0.55\le x\le 0.6$) possess multiple ZBPs of tunneling in the vortex cores~\cite{2018arXiv181208995M}. Here, we incorporate this experimental data of the scanning tunneling microscope~(STM) and the simulation of the Majorana surface physics to show the evidence that an s-wave superconducting surface with a single Dirac cone, as an effective 2D $p \pm ip$ superconductor (SC), can host multiple Majorana zero modes~\cite{Fu2008Superconducting}, which can form scalable Majorana qubits. This iron-based superconductor can potentially be an ideal platform for Majorana quantum computing~\cite{RMP_braiding} and interacting Majorana systems~\cite{Chiu_strong_inter_Majorana}. 

Since in the presence of the vortex the superconductor order parameter phase winding aligns the spin rotation texture in the surface Dirac cone, a Majorana vortex mode (MVM) with zero energy (Majorana zero mode) emerges in the vortex core. The first candidate of this effective $p \pm ip$ superconductor is the heterostructure of a topological insulator (${\rm Bi}_2{\rm Te}_3$) and an s-wave superconductor (${\rm NbSe}_2$). 
Recent scanning tunneling spectroscopy measurement observed strong zero-bias conductance peaks at the vortex cores of the heterostructure~\cite{Xu2015Experimental, Sun2016Majorana}, consistent with MVM interpretation in tunnel spectroscopy~\cite{Law2009Majorana}. However, the measurement resolution is not high enough to rule out other low-energy modes, Caroli-de Gennes-Matricon (CdGM) modes~\cite{CAROLI1964307}. The two different modes residing in the vortex cores had been indistinguishable until revealing the topology of the type-II iron-based topological superconductor (${\rm FeTe}_{x}{\rm Se}_{1-x}$)~\cite{10.1093/nsr/nwy142}.  This material possesses bulk Fermi surfaces and Dirac-cone-type spin-helical surface states, for which SC pairing can be induced by the bulk $s$-wave superconductivity~\cite{Zhang2018Observation}. The Fermi energies of the bulk and surface states are small enough to lead to the energy of the CdGM modes being much greater than the STM resolution. Hence, while the tunneling peaks away from zero bias voltage have been observed and indicated the CdGM modes~\cite{Chen:2018aa}, the tunneling observation of the ZBPs in the vortex cores~\cite{Wang2018Evidence} has provided the first clue of MVM existence. The further hint of the MVM existence is supported by the recent observation of the tunneling conductance plateau that is \emph{near} the quantized value~\cite{2019arXiv190406124Z,Chen_2019}. 

\begin{figure}[t]
\begin{center}
\includegraphics[clip,width=0.7\columnwidth]{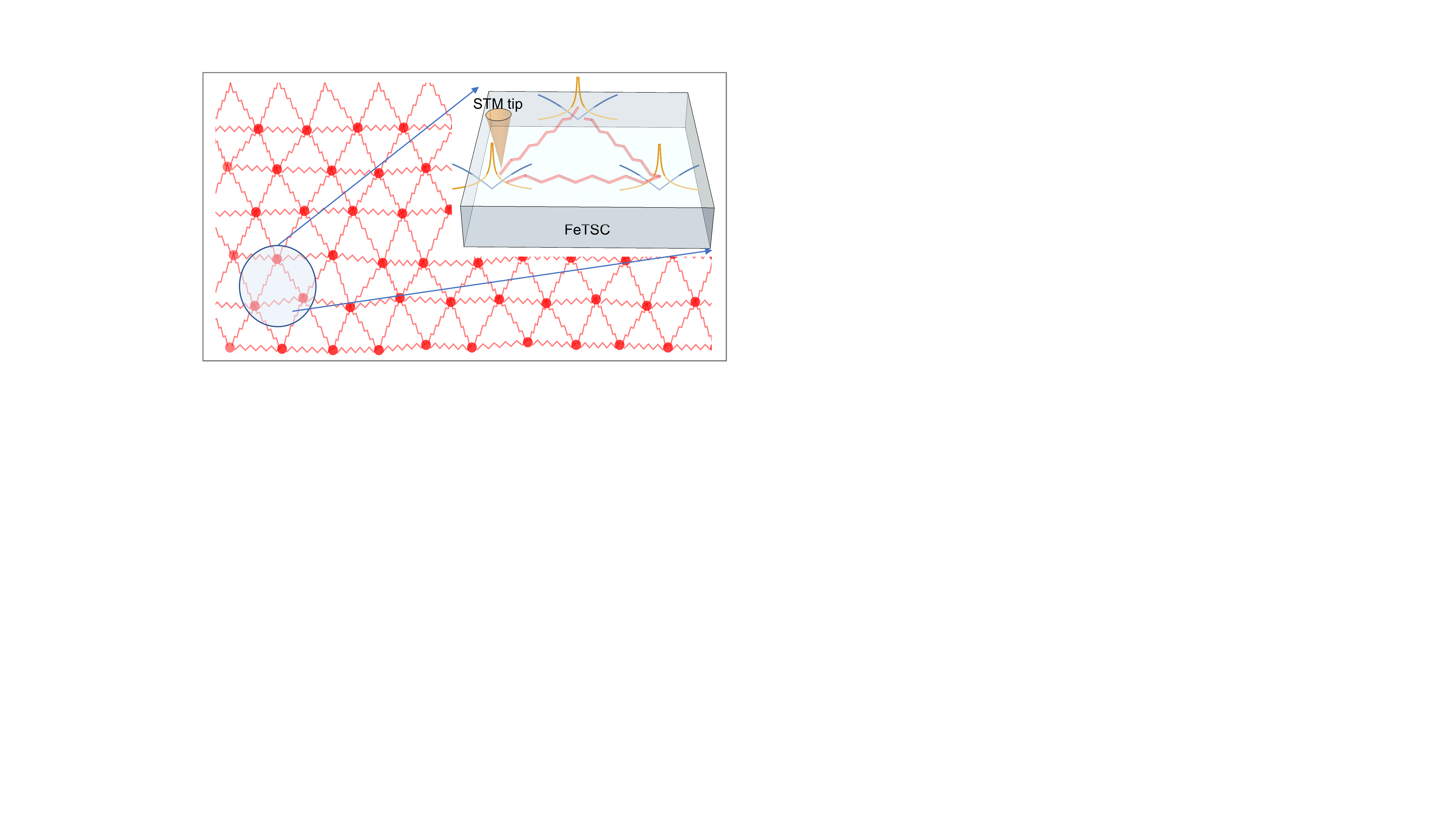}
\caption{  {\bf  Illustration of low-energy physics of Majorana hybridizations in disordered vortex lattice. } The wiggle line links two nearest neighbor MVMs (red dots). The inset shows the vortices on the surface of FeTe$_{x}$Se$_{1-x}$ probed by the STM tip, where the dark yellow curves represent in the MVM distributions and the blue curves the superconductor gaps near the vortex cores. 
 }
\label{main_physics} 
\end{center}
\end{figure}

Recently, the high-energy resolution STM~\cite{doi:10.1063/1.5049619} probed hundreds of vortex cores on the ($001$) surface of ${\rm FeTe}_{x}{\rm Se}_{1-x}$ in the different magnetic fields. The experimental data show that the fraction $R_{\rm{ZBP}}$ of vortex cores possessing the ZBPs of tunneling spectra monotonically decreases as the magnetic field increases~\cite{2018arXiv181208995M}. 
Another experimental group~\cite{2019arXiv190901686C} independently has observed the similar decreasing behavior of the ZBP rate, although the rate  $R_{\rm{ZBP}}$ is much lower than the previous group operating the higher energy solution of the STM.
Although as illustrated in Fig.~\ref{main_physics} the Majorana hybridization might be a straightforward explanation, this picture cannot directly lead to the monotonically decreasing rate $R_{\rm{ZBP}}$ of the observed ZBPs without considering other factors.  The reason is that the energy splitting of the two-Majorana hybridization exhibits an oscillation with the exponential decay as the intervortex distance increases~\cite{PhysRevLett.103.107001,:2010tb}. In this regard, the ZBP rate $R_{\rm{ZBP}}$ should have an oscillating behavior, which has not been observed in ${\rm FeTe}_{x}{\rm Se}_{1-x}$.

Instead of the focus on the physics of individual vortex, the main goal of this manuscript is to investigate whether in all the observed vortex cores the global features of the ZBPs are linked to the nature of multiple MVMs. We build a 3D Bogoliubov-de Gennes tight-binding (TB) model capturing the effective low-energy physics ($E<0.2$meV) of over a hundred MVMs on the (001) surface of ${\rm FeTe}_{x}{\rm Se}_{1-x}$; the TB model includes only one surface Dirac cone with an s-wave superconducting pairing, which leads to the presence of MVMs. In this regard, we simplify the TB model in the second quantization form of  
\begin{align}
\hat{H}_{\rm{BdG}}=\hat{H}_{\rm{TI}}+ \sum_\br \Big [\Delta (\br)C_{\br \downarrow}^\dagger  C_{\br \uparrow }^\dagger- \Delta (\br)C_{\br \uparrow}^\dagger  C_{\br \downarrow }^\dagger + h.c. \Big ]. \label{TB first}
\end{align}
and we construct the Hamiltonian of the topological insulator $\hat{H}_{\rm{TI}}$ to possess a surface Dirac cone being identical to the one in ${\rm FeTe}_{x}{\rm Se}_{1-x}$~\cite{Zhang2018Observation,Wang2018Evidence}. Moreover, we insert Abrikosov vortices by adding phase windings in the order parameter and vector potential~\cite{Liu_2015,Chiu_strong_inter_Majorana} in the hopping terms (see the explicit form of the TB model in Supplementary Materials). It is known that the self-consistent theory can determine the values of the superconductor parameters. Circumventing the subtle self-consistent electronic structure of multiple vortices, we directly adapt the values of the physical parameters from the experimental data~\cite{Zhang2018Observation,Wang2018Evidence,PhysRevB.95.224505,PhysRevB.81.180503} for the TB model.  


Our key finding is that the Majorana hybridization and the disordered vortex distribution result in the decreasing ZBP rate $R_{\rm{ZBP}}$ with the increasing magnetic field. Moreover, by analyzing the experimental data~\cite{2018arXiv181208995M} we find the statistics of the observed energy peaks agrees with the one from the simulation including the physics of the Majorana hybridization. That is, the agreement between the analyzed experimental data and the Majorana simulation suggests the existence of the scalable MVMs. In addition, we study the interplay of the vortex location and the ZBPs in the vortices to show that the subtle relation between the energy spectra and the Majorana couplings is consistent with our new experimental observation. To confirm multiple MVMs in the iron-based superconductors, we propose spin-selected STM for the measurement of the universal Majorana spin signature. With further experimental confirmation in the future, the surface of this iron-based superconductor can be a platform hosting scalable MVMs for quantum computing.  






\vspace{0.5cm}


\noindent
\textbf{\large Results} \\
\noindent
{\bf Majorana Physics on the surface of ${\rm FeTe}_{x}{\rm Se}_{1-x}$}

We start with one single vortex through the iron-based superconductor FeTe$_{x}$Se$_{1-x}$.  The iron-based superconductor possesses a bulk-band inversion gap so that a Dirac cone protected by time-reversal symmetry is present on its surface. Furthermore, the recent ARPES measurement \cite{Zhang2018Observation} has shown that  an s-wave superconducting gap is induced in the surface Dirac cone. Hence, this superconducting Dirac cone surface state leads to the non-trivial topology~\cite{Fu2008Superconducting}. Since our focus is on only the low-energy physics of FeTe$_{0.55}$Se$_{0.45}$, the effective physics of the topological superconductor can be described by a 2D Bogoliubov-de-Gennes (BdG) Hamiltonian of the surface Dirac cone with an s-wave superconducting pairing
\bee
H_{\rm{BdG}}^{\rm{surf}}(\bk)=\bma 
-\mu & \nu_F (k_x-ik_y)   & 0 & -\Delta \\
\nu_F (k_x+ik_y) & -\mu & \Delta & 0 \\
0 & \Delta^*  & \mu  & \nu_F (k_x+ik_y) \\
-\Delta^*  & 0 &  \nu_F (k_x-ik_y)  & \mu \\
\ema, \label{continuous model}
\ee
in the basis of $C_\bk=(c_{\uparrow  \bk}^{},c_{\downarrow  \bk}^{},c_{\uparrow  \bk}^\dagger,c_{\downarrow  \bk}^\dagger)$, where $\mu$ is the chemical potential, $\nu_F$ the Fermi velocity, and $\Delta_0$ the superconducting gap. To reveal the topological superconductivity, we introduce a vortex at $\br=0$ so that the order parameter has the additional winding phase $\Delta=\Delta_0 e^{i\theta}$. 
\begin{widetext}
Hence, with broken translation symmetry, since the momentum $\bm{k}$ is not a good quantum number, the surface Hamiltonian has to be rewritten in real space 
\bee
H_{\rm{BdG}}^{\rm{surf}}(\br)=\bma 
-\mu & -i\nu_F e^{-i\theta}(\partial_r-\frac{i}{r}\partial_\theta)  & 0 & -\Delta_0 e^{i\theta} \\
-i\nu_F e^{i\theta}(\partial_r+\frac{i}{r}\partial_\theta) & -\mu & \Delta_0 e^{i\theta} & 0 \\
0 & \Delta_0 e^{-i\theta} & \mu  & -i\nu_F e^{-i\theta}(\partial_r+\frac{i}{r}\partial_\theta) \\
-\Delta_0 e^{-i\theta} & 0 & -i\nu_F e^{-i\theta}(\partial_r-\frac{i}{r}\partial_\theta) & \mu \\
\ema, \label{cont model}
\ee
\end{widetext}
in the basis of $C_\br=(c_{\uparrow  \br}^{},c_{\downarrow  \br}^{},c_{\uparrow  \br}^\dagger,c_{\downarrow  \br}^\dagger)$. By solving the eigenvalue problem, the wavefunction of the MVM with zero energy trapped in the vortex is in the form of \cite{PhysRevLett.115.177001,PhysRevB.94.224501}
\bee
\ket{{\rm{MVM}}}=e^{-r/\xi}(J_0(k_F r),J_1(k_F r)e^{i\theta},J_0(k_F r),J_1(k_F r)e^{-i\theta})^T, \label{cont state}
\ee
where $J_i(k_Fr)$ is the Bessel function of the first kind, the coherence length $\xi=\nu_F/\Delta_0$, and the Fermi momentum of the surface Dirac cone $k_F=\mu/\nu_F$. The coherence length $\xi$ represents the spatial size of the MVM 
and the Fermi wavelength $1/k_F$ represents the oscillation length of the MVM due to the oscillation of the Bessel functions. We note that the Majorana coherence length is physically different from the superconducting coherence length, which represents the length scale of the vanishing superconducting gap in a vortex, although these two lengths are close in the BCS theory.

Because these two key lengths characterize the MVM, experimentally determining these characteristic lengths is essential to confirm the existence of the MVMs. For ${\rm FeTe}_{x}{\rm Se}_{1-x}$, we adapt superconductor gap $\Delta_0=1.8$meV, Fermi velocity $\nu_F=25$nm$\cdot$meV, and Fermi momentum $k_F=0.2$nm$^{-1}$ based on the recent experimental measurements of angle-resolved photoemission spectroscopy~\cite{Zhang2018Observation,Wang2018Evidence}; hence, the values of the two characteristic lengths are given by the coherence length $\xi=13.9$nm and the Fermi wavelength $1/k_F=5$nm. Using these characteristic lengths, we build a three-dimensional TB for the simulation and Fig.~\ref{one_vortex_up_down}(a,b) shows the wavefunction of the single MVM in the continuum model (\ref{cont model}) and the TB model are in perfect agreement.   


\begin{figure}[t]
\begin{center}
\includegraphics[clip,width=0.99\columnwidth]{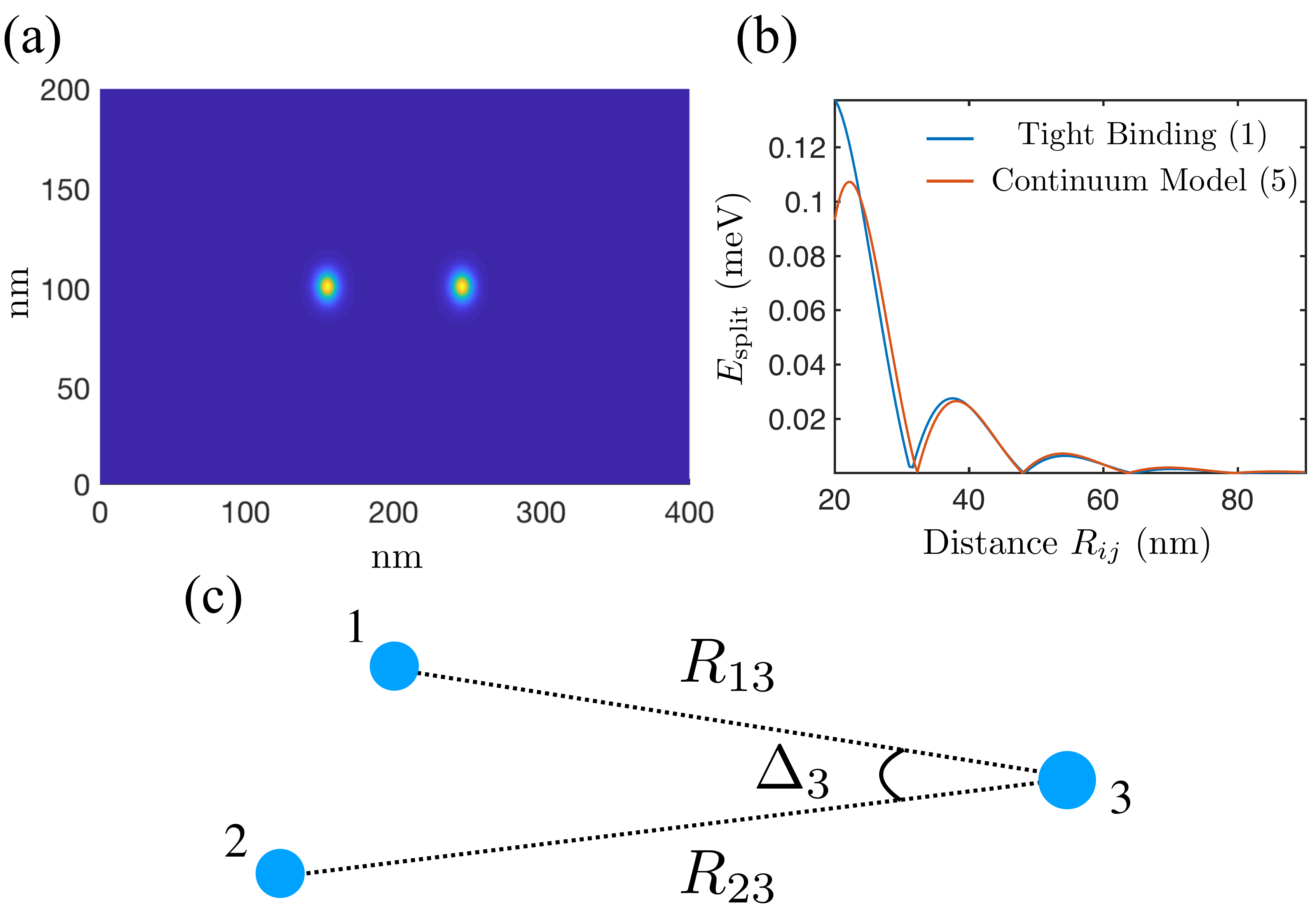}
\caption{  {\bf The hybridization of the two MVMs on the surface of the $400\times 200\times 10$nm$^3$ FeTe$_{x}$Se$_{1-x}$ and the hybridization strength is also affected by the presence of other vortices.} (a) shows the distribution of the two MVMs on the top surface of the system from the TB model (\ref{TB first}). (b) The hybridization energy of the two MVMs as a function of the distance $R_{ij}$ in Eq.~\ref{splitting} is in agreement with the TB model. (c) Vortex (MVM) 3 can affect the hybridization strength of Majoranas 1 and 2. The main factor stems from the phase difference $\Delta_3$ in the hybridized Majoranas. 
 }
\label{Two_Majorana} 
\end{center}
\end{figure}

\vspace{0.5cm}

\noindent
{\bf Hybridization of two Majorana vortex modes}

	The hybridization of the two MVMs plays an important role to affect the presence of the ZBPs in the vortex cores. By considering two vortices with distance $R_{ij}$, we have two MVMs residing in the two vortex cores respectively as illustrated in Fig.~\ref{Two_Majorana}(a). The energy hybridization of the two MVMs is captured by a distance $R_{ij}$ dependent term and a phase term affected by all of the vortices separately ($t_{ij}\propto t_{R}t_{\rm{phase}} $). For $k_Fr\gg 1$, due to the asymptotic form of the Bessel function, the distance-dependent hybridization energy of the two MVMs in the continuum model is given by~\cite{PhysRevLett.103.107001,:2010tb} 
\begin{align}\label{splitting}
t_{ij}\propto t_{R} =  \frac{\cos(k_F R_{ij}+\theta)}{\sqrt{R_{ij}}}e^{-R_{ij}/\xi}.
\end{align}
Although this hybridization approximation holds in the large $k_F r$ limit, as shown in Fig.~\ref{Two_Majorana}(b) this is in good agreement with the TB model, which naturally includes the Majorana hybridization. (See Supplementary Materials for the details of the TB model.) It is known that the phase factor in the continuum model is given by $\theta=\pi/4$; however, in reality, $\theta$ depends on the details of the system. For example, for the TB model of the two vortices $\theta=1.40$ in our simulation for FeTe$_{0.55}$Se$_{0.45}$. In the case of the iron-based superconductor, the Majorana hybridization exhibits the oscillation with the length period $\pi/k_F=15.7$nm. On the one hand, in the STM experiment~\cite{2018arXiv181208995M}, the intervortex distance is varied from $46.5$nm to $19.2$nm as the magnetic field is increased from $1$T to $6$T. Since the oscillation length of the Majorana hybridization is within the experimental range, the ZBP rate $R_{\rm{ZBP}}$ controlled by the hybridization should be somehow oscillating as the magnetic field increases. 
However, the oscillation has never been observed.
The main goal of this manuscript is to resolve this puzzle by considering additional factors in Majorana physics. 


	The hybridization energy is also affected by the phases from all the vortices on the surface. We introduce the third vortex away from those two hybridized MVMs as an example in Fig.~\ref{Two_Majorana}(d); the idea can be easily extended to multiple vortices. While the third vortex contributes additional superconducting gap phases to the first two MVMs~\cite{PhysRevLett.111.136401}, $h/2e$ magnetic flux with London penetration depth ($\lambda \sim 500$nm~\cite{PhysRevB.81.180503,PhysRevB.82.184506}) goes through each vortex and the vector potential of the magnetic flux from each vortex affects the strength of the Majorana hybridization~\cite{Chiu_strong_inter_Majorana,Liu_2015}. Apart from the oscillation and exponential decay (\ref{splitting}), the Majorana hybridization energy is proportional to an additional term~\cite{Liu_2015,Chiu_strong_inter_Majorana} 
\begin{equation}\label{phase coupling}
t_{12} \propto t_{\rm{phase}}=\sin \omega_{12}, \ {\rm{where}}\ \omega_{12}= \int_{\bm{r}_1}^{\bm{r}_2} \Big ( \frac{1}{2}\nabla \theta - \frac{e}{\hbar} \bm{A} \Big ) \cdot d\bm{l}, 
\end{equation}
where $\bm{A}$ is the vector potential stemming from the magnetic flux and $\int_{\bm{r}_i}^{\bm{r}_j} \nabla \theta \cdot d\bm{l}=\theta_j - \theta_i$ is the phase difference between the superconducting gap at the first two hybridized MVMs. This coupling relation is directly included in our TB model with the magnetic flux. For few vortices residing (three in Fig.~\ref{Two_Majorana}(d)) in a small superconductor~\cite{SERRIERGARCIA2017109} when all of the intervortex distances are much less than the London penetration depth $\lambda$, the magnetic flux is diluted so that the vector potential $\bm{A}$ can be neglected (see Eq.~\ref{vector potential all}). Hence, $\omega_{12}=(\pi+\Delta_3)/2$, where $\pi$-phase stems from the first two vortices and $\Delta_3$ is the phase difference at the two coupling vortices from the third vortex as shown in Fig.~~\ref{Two_Majorana}(d). When the third vortex is moved away from the first two vortices so that the distances are much greater than the London penetration depth ($R_{13},\ R_{23}\gg \lambda$), $\Delta_3$ becomes small and the magnetic flux from the third vortex screens out the contribution of $\omega_{12}$ from the third vortex 
\bee
(\frac{1}{2} \nabla \theta - \frac{e}{\hbar} \bm{A})_3 \approx 0. \label{screening}
\ee
That is, the far-away vortex no longer affects the hybridization of the two Majoranas. This gives rise to the minimal simulation size of the TB model, which must be greater than the London penetration depth. 

\begin{figure}[t]
\begin{center}
\includegraphics[clip,width=0.99\columnwidth]{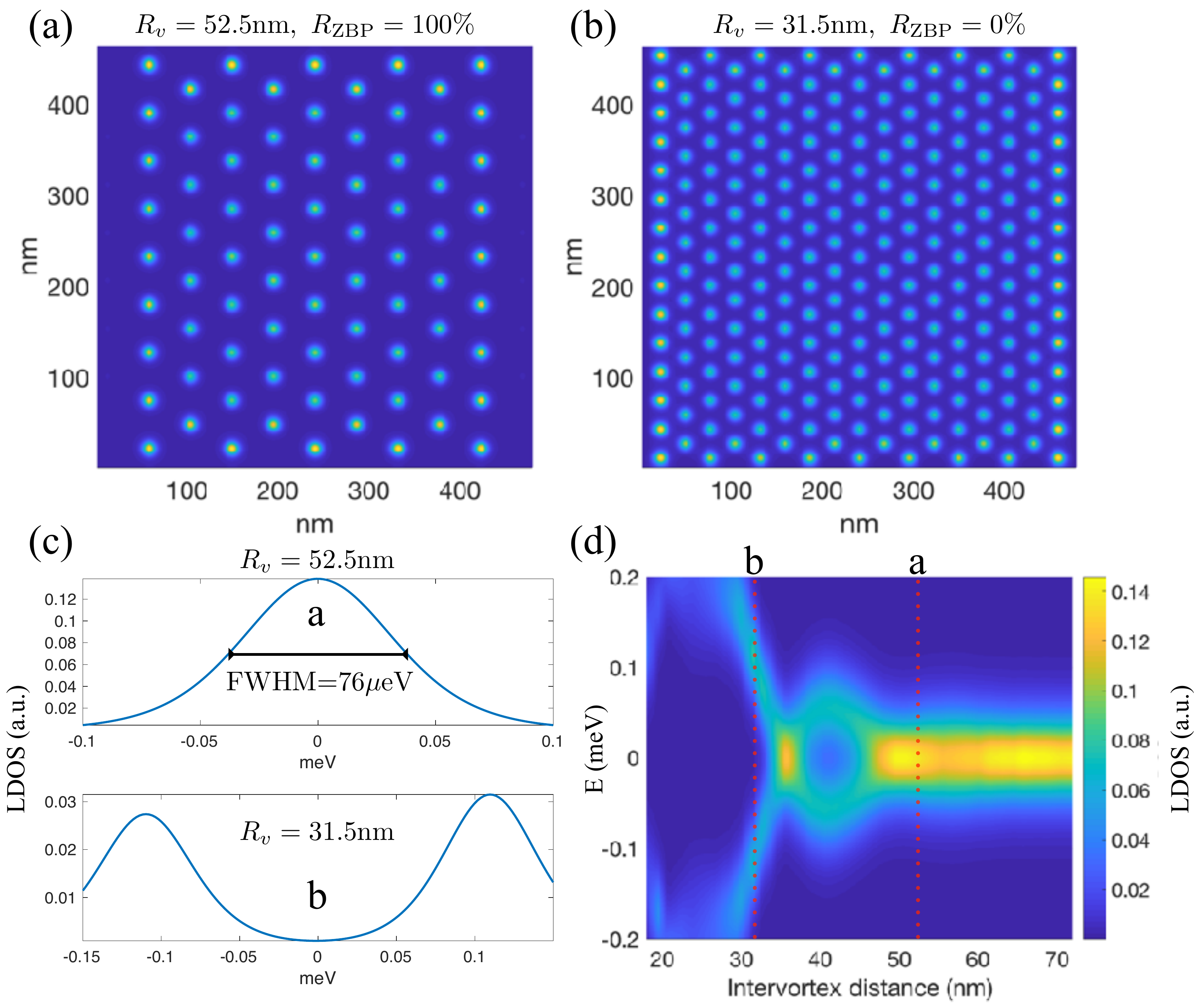}
\caption{   {\bf Triangular vortex lattice in FeTe$_{x}$Se$_{1-x}$ tight binding model.} (a,b) show LDOS with energy within $\pm 12\mu$eV and $\pm 0.12$meV respectively and the vortex lattice distributions with different intervortex distances $R_v$ on the top surface. All of the vortex cores $100$nm away from the boundary have identical LDOS. (c) the LDOS in the vortex cores shows a ZBP for $R_v=52.5$nm and broken particle-hole symmetry for $R_v=31.5$nm as the two peak heights are different. (d) the LDOS in the vortex cores shows that the locations of the energy peaks as a function of $R_v$ exhibit the oscillation and decay being consistent with the hybridization of the two MVMs (\ref{splitting}). 
 }
\label{triangular_lattice} 
\end{center}
\end{figure}



\vspace{0.5cm}

\noindent
{\bf Majorana vortex triangular lattice}




\begin{figure*}[t]
\begin{center}
\includegraphics[clip,width=1.70\columnwidth]{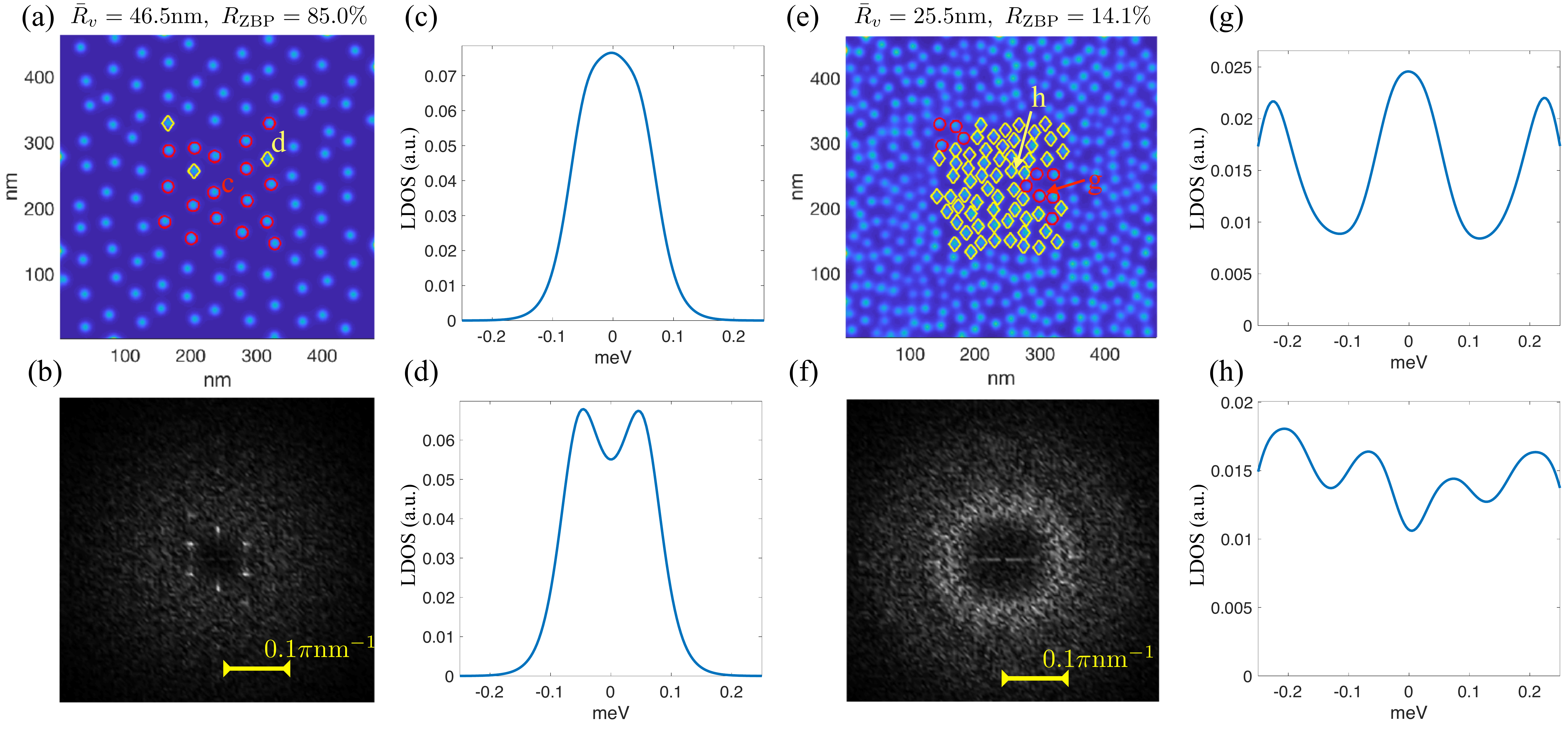}
\caption{ {\bf Distorted vortex lattice and glass vortex distribution based on the STM experiment~\cite{2018arXiv181208995M}} (a,e) show LDOS with energy within $\pm 30\mu$eV and $\pm 0.14$meV respectively. (a,b) show distorted vortex triangular lattice in real and Fourier spaces respectively with $\bar{R}_v=46.5$nm ($B\approx 1$T). (c,d) the LDOS of the vortex cores with/without ZBPs exhibit one/two peaks. (e,f) show glass vortex distribution in real and Fourier spaces respectively with $\bar{R}_v=25.5$nm ($B\approx 3.33$T). (g,h) the LDOS of the vortex cores with/without ZBPs exhibits multiple peaks. 
 }
\label{ZBP_distribution} 
\end{center}
\end{figure*}

	We extend the system from few vortices to triangular vortex lattice on the surface of FeTe$_{x}$Se$_{1-x}$. The vector potential $\bm{A}$ from the penetrating magnetic field cannot be neglected since the magnetic flux is accumulated in the presence of the multiple vortices in the wide spatial range.  The simple continuum model (\ref{continuous model}) of the single vortex is unable to capture the surface physics of the vortex lattice including the magnetic flux. Therefore, our TB model provides a straightforward way to reveal the Majorana lattice physics and the mechanism of the TB model naturally includes the physics of the Majorana hybridization. There are two important characteristic lengths for the simulation --- Majorana coherence length $\Delta/\nu=13.9$nm and London penetration depth $\lambda\sim 500$nm. For the simulation system, not only the surface size must be much greater than the Majorana coherence length but also the virtual surface size must be much greater than the London penetration depth. (See sec.~\ref{simulation details} in Supplementary Materials for the details) Namely, the virtual size means that virtual vortices, which are present outside the physical simulation system, with the magnetic flux and their SC phase windings are added in the physical system. 
	Since $\lambda \gg  \Delta/\nu$ for FeTe$_{x}$Se$_{1-x}$, the real surface size for the simulation can be much smaller than the virtual surface size. By using this simulation setup, the MVMs in the central region of the system do not encounter the finite size effect, since the coupling between any two MVMs in the center and on the boundary is suppressed due to the exponential decay of the Majorana wavefunction (\ref{cont state}).  Furthermore, by Eq.~\ref{screening} the contribution of the phase factor $\omega_{ij}$ from vortices outside the virtual system has been screened out. 

	The distributions of the triangular vortex lattice with intervortex distance $52.5$nm and $31.5$nm are shown in Fig. \ref{triangular_lattice}(a,b). Each vortex core in the central region of the surface possesses the identical local density of states (LDOS) with fixed $R_v$ as an effective lattice translational symmetry due to the screening effect from the magnetic flux.  In the absence of the magnetic flux, the translation symmetry in the central region is completely broken as shown in Fig.~\ref{lpdinf} since the virtual surface size for the simulation always is not big enough to reach the screening boundary as the effective London penetration depth $\lambda\rightarrow \infty$. Hence, introducing the magnetic flux plays an important role to depict the real experimental systems with finite London penetration depth. 

	Choosing effective temperature $T$ for the simulation is also essential to capture the observed LDOS of the vortex cores. Here we assume the tunneling spectra measured by STM are identical to the LDOS from TB model simulation. In the experiment the average full width at the half maximum (FWHM) of the LDOS peak in the low field ($B=1T$, $R_v=46.5$nm) is around 80 $\mu$eV as shown in Fig.~\ref{supp_FWHM}. Hence, we choose $T=20\mu$eV so that as shown in Fig.~\ref{triangular_lattice}(c)($R_v=52.5$nm) the FWHM of the ZBPs in the TB model is given by $76\mu$eV, which is reasonably close to the experimental value~\cite{2018arXiv181208995M} (see Fig.~\ref{supp_FWHM} in Supplementary Materials). Since each vortex core possesses the identical LDOS with fixed $R_v$, Fig.~\ref{triangular_lattice}(d) shows the LDOS as a function of the intervortex distance $R_v$. The ZBPs persist for $R_v\ge 48$nm since the temperature broadening erases the effect of the small energy splitting. The FWHM of the ZBPs is always close to $80\mu$eV for different $R_v$'s in this region.

	For $R_v< 48$nm, the oscillation of the peak location shares the similarity with the hybridization energy (\ref{splitting}) of the two MVMs as a function of the intervortex distance as shown in Fig.~\ref{Two_Majorana}(c). When the MVMs get closer, the LDOS starts to break particle-hole symmetry due to the Majorana hybridization and wavefunction overlapping (see sec.~\ref{PH symmetry broken} in Supplementary Materials for the detailed reason). Fig.~\ref{triangular_lattice}(c)($R_v=31.5$nm) shows different peak heights with the opposite energies. On the other hand, another ZBP appears at $R_v\approx 36$nm. 

	The lattice translational symmetry leads to either $100\%$ or $0\%$ ZBP rate $R_{\rm{ZBP}}$ at the vortex cores. Therefore, the triangular vortex lattice cannot explain the monotonically decreasing \emph{fractional} ZBP rate. This leads to further investigation of the vortex distribution on the surface.

 \vspace{0.5cm}
 
\noindent
{\bf Disordered Majorana vortex distribution}	\\
In reality, due to the imperfection and impurity of the crystal, the distribution of the vortices in ${\rm FeTe}_{x}{\rm Se}_{1-x}$ is not perfect triangular lattice anymore. The experimental observation~\cite{2018arXiv181208995M} shows in the Fourier space the vortex distributions resemble distorted triangular lattice in the low magnetic field and glass distribution in the high field. For the simulation, we set up these two types of vortex distributions in the TB model for different average intervortex distances ($18\le \bar{R}_v\le 48$nm cf.~the experimental range $19\sim46.5$nm~\cite{2018arXiv181208995M}). 
We examine if on the top surface LDOS in each vortex core within the central region ($100\times100$nm) possesses a ZBP. First, we examine the Majorana physics in the low field ($B=1T$) corresponding to the average intervortex distance $\bar{R}_v=46.5$nm in the experiment. Fig.~\ref{ZBP_distribution}(a,b) show the vortex distribution of the distorted triangular lattice in real and Fourier spaces respectively. Due to the translational symmetry breaking, the ZBP rate $R_{\rm{ZBP}}=85.0\%$ is fractional while red circles and yellow diamonds indicate vortex cores with/without ZBPs in Fig.~\ref{ZBP_distribution}(a). The LDOS of the selected vortex cores with/without ZBPs is shown in Fig.~\ref{ZBP_distribution}(c,d) respectively. On the other hand, in the high field $B=3.33T$, which corresponds to short intervortex distance ($\bar{R}_v=25.5$nm), the vortices distribute like glass, which has a ring-shaped distribution~\cite{RAMAN:1927aa} in Fourier space, as shown in Fig.~\ref{ZBP_distribution}(e,f) in the Fourier and real spaces respectively. The ZBP rate  $R_{\rm{ZBP}}=14.1\%$ becomes much smaller by comparing with the low magnetic field (a). It is worth to notice that the LDOS of the selected vortex cores with/without ZBPs in Fig.~\ref{ZBP_distribution}(g,h) possess multiple peaks. Since the energies ($\ge 0.96$meV) of the CdGM modes are much higher than the energy peaks ($\le 0.2$meV) of the LDOS, the multiple peaks stem from the hybridization of the multiple MVMs, instead of CdGM modes. (See sec.~\ref{Simulation} in Supplementary Materials for the CdGM energies.)

	
	
\begin{figure}[t!]
\begin{center}
\includegraphics[clip,width=0.99\columnwidth]{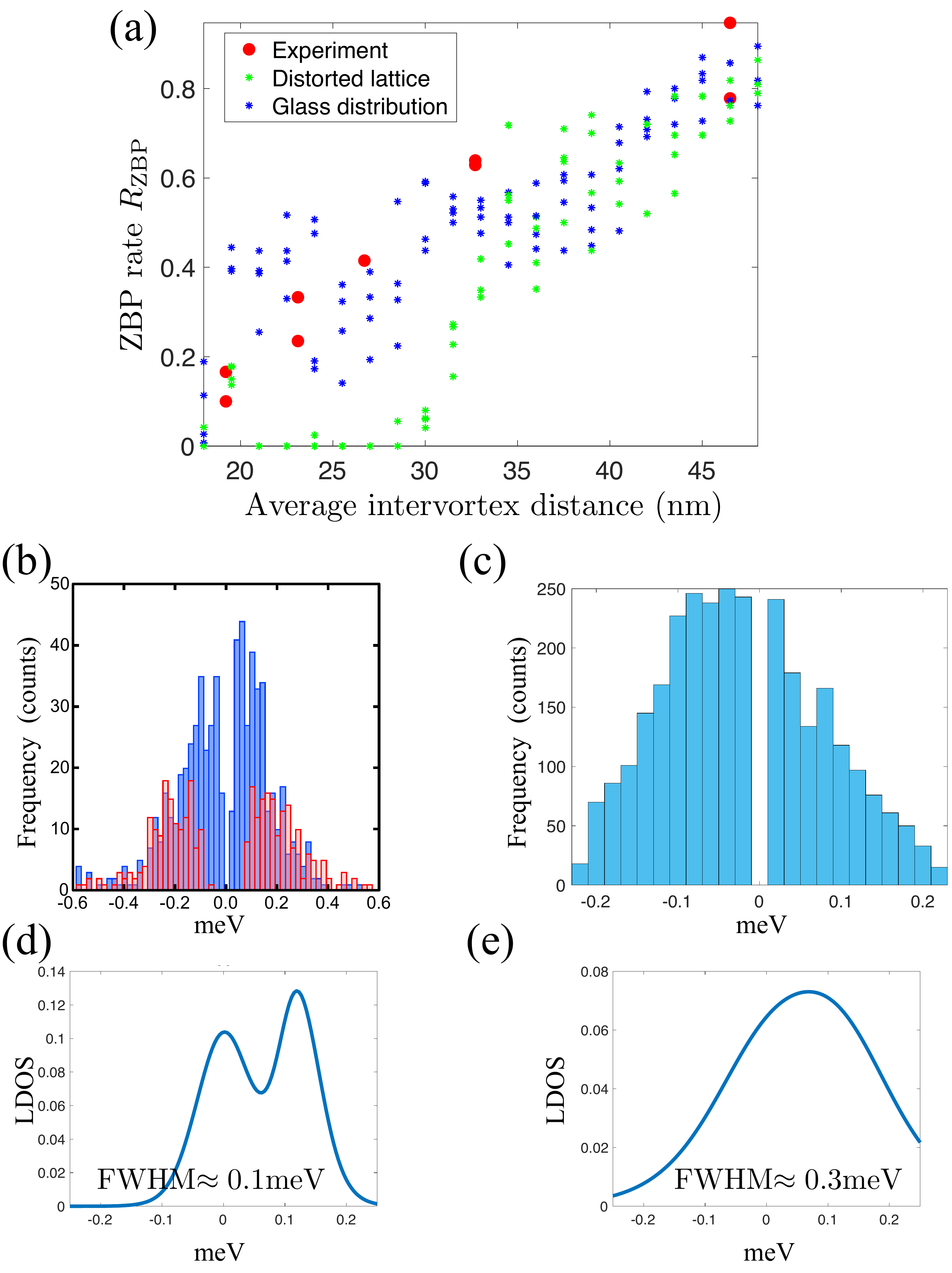}
\caption{ {\bf LDOS peak statistics} (a) the ZBP rates $R_{\rm{ZBP}}$ for the two different types of the vortex distributions. The experimental ZBP rate (red dot) is consistent with distorted lattice (green) in the low magnetic field and glass distribution (blue) in the high field. (b) the frequency of the first peaks (blue) closest to zero energy in the vortex cores without ZBPs from the experimental data \cite{2018arXiv181208995M} has pyramid-shaped distribution (except for zero energy).  The red bars indicate the frequency of the second peaks closest to zero energy in the vortex cores with ZBPs. The second peaks exhibit the wide range of the energy spreading and are located at least $\pm 60\mu$eV away from zero energy. (c) the frequency of the first peaks (blue) closest to zero energy from the TB model share the similar distribution with the experimental data. (d,e) represent the LDOS of the same vortex core with different peak broadenings. The FWHMs for a single peak are given by $0.1$meV and $0.3$meV respectively. 
 }
 \label{ZBP_height_and_rate} 
\end{center}
\end{figure}

	We consider the two different types of vortex distributions for the TB model simulation. Our key result presenting in Fig.~\ref{ZBP_height_and_rate}(a) shows overall the ZBP rate $R_{\rm{ZBP}}$ in the TB model qualitatively decreases when the intervortex distance becomes shorter with the increasing magnetic field. We note that some of the physical parameters (e.g.~$\Delta$) may vary as the magnetic field is changed. Since the upper critical magnetic field is close to $50$T~\cite{PhysRevB.81.184511}, we assume that the magnetic field strength ($1\sim 6$T) does not affect the physical parameters, except for the average intervortex distance. In the low magnetic field ($\bar{R}_v\ge 33$nm) the ZBP rates of the two different vortex distributions share the similar decay consistent with the experimental observations, while the vortex distribution transits from the distorted lattice to glass distribution in the high magnetic field region. Since the ZBP statistics from the simulation does not exhibit any oscillation as a function of the intervortex distance, losing the perfect triangular lattice distribution of the vortices can be a primary factor smearing the oscillation feature of the Majorana coupling. The stronger the average strength of the Majorana coupling is, the shorter the average intervortex distance is. The stronger Majorana coupling increases the chance of the MVMs away from zero energy. Therefore, the disordered vortex distribution and the hybridization of the MVMs are the keys leading to the monotonic decreasing ZBP rate. In other words, the observation of the decreasing ZBP rate provides a clue to the multiple Majorana hybridizations on the surface of ${\rm FeTe}_{x}{\rm Se}_{1-x}$. 
	We further examine if inhomogeneous superconducting gaps or inhomogeneous chemical potentials can affect the LDOS of the vortex cores. The inhomogeneity of the superconducting gaps does not alter the LDOS in the vortex cores, whereas non-uniform chemical potentials can affect the LDOS (See sec.~\ref{disorder_cp} in Supplementary Materials for the details). However, because Se and Te are isovalent, the spatial inhomogeneity of the chemical potential should be very small~\cite{PhysRevB.98.144519}. Only the vortex distribution is the main physical factor affecting the LDOS of the vortex cores.

 \vspace{0.5cm}


\begin{figure*}[thb]
\begin{center}
\includegraphics[clip,width=1.90\columnwidth]{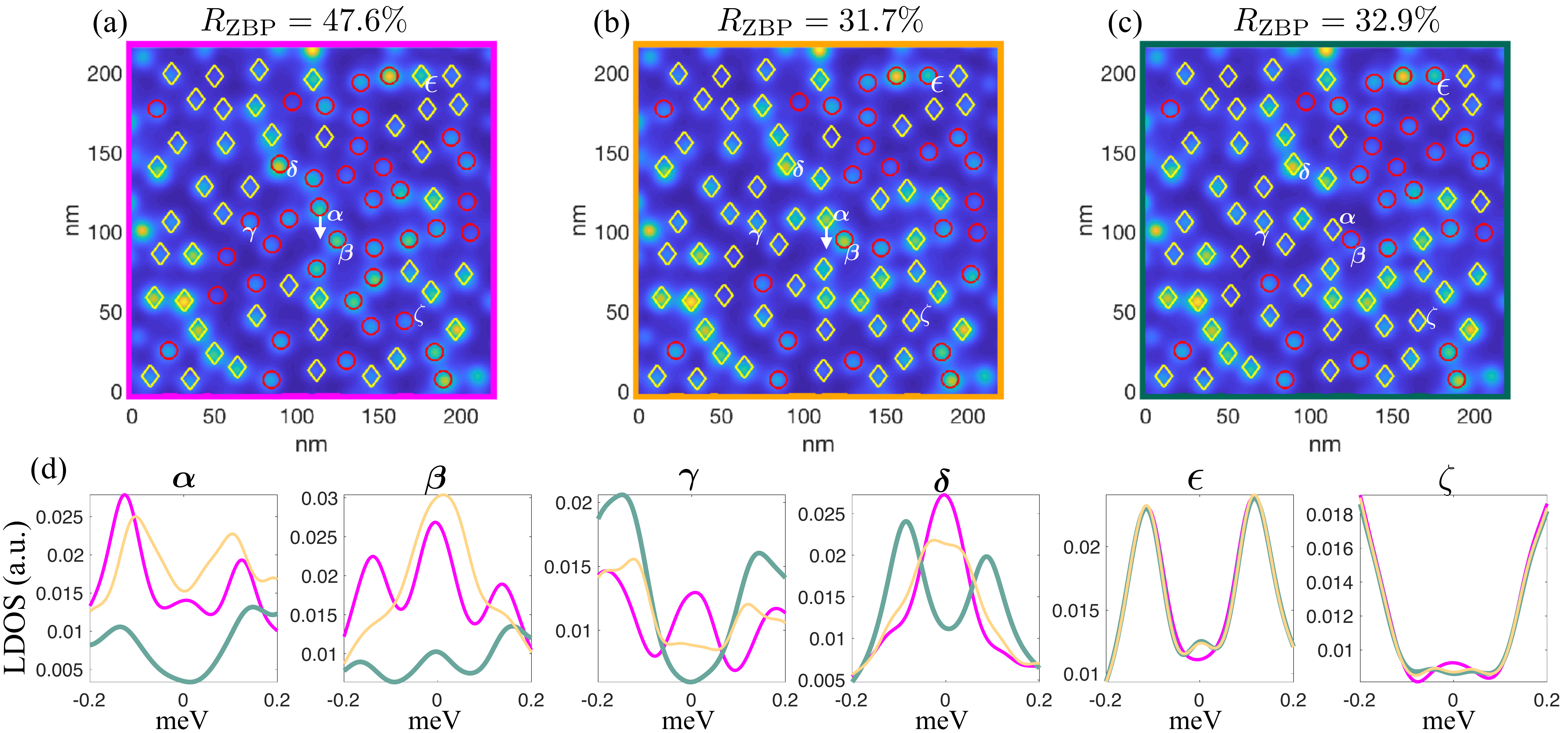}
\caption{ {\bf The comparison of the LDOS before and after the movement of the single vortex. The movement of the single vortex significantly changes the LDOS of the vortex cores in the wide range. } (a,b,c) show LDOS with energy within $\pm 0.13$meV.  With $24$nm average intervortex distance ($B\approx 4$T) the vortex distributions are almost identical, except for the vortex $\alpha$ in the center. The red circles and yellow diamonds indicate the vortex cores with/without ZBPs respectively. (a) at the beginning, the central vortex $\alpha$ possesses a ZBP. (b) after moving $7$nm away from the original point, the central vortex $\alpha$ does not have the ZBP. (c) shows another $7$nm movement of the central vortex. (d) shows LDOS (a:pink, b:orange, c:green) for different vortex cores. Near the moving vortex the LDOS of the vortex cores ($\alpha,\beta,\gamma,\delta$) has significant changes, while away from the moving vortex the LDOS ($\epsilon,\zeta$) exhibits slight changes. 
 }
\label{vortex_move} 
\end{center}
\end{figure*}

\noindent
{\bf Statistics of energy peaks}	\\
Examining the statistics of the first peak in each vortex core closest to zero energy in the experimental data \cite{2018arXiv181208995M}, we find another important clue that supports the surface physics stemming from the hybridization of multiple MVMs. 
By excluding the vortex cores possessing ZBPs, most of the first peaks in the remaining vortex cores as shown in fig.~\ref{ZBP_height_and_rate}(b)(blue bars) are very close to zero energy.  We also use the simulation result for the Majorana surface physics to plot the number distribution of the first peaks in fig.~\ref{ZBP_height_and_rate}(c). The experimental data and simulation share a similar peak distribution showing the number of the peaks decreases as the energy moves away from zero energy. In this regard, these peaks slightly away from zero energy can be explained by the hybridization of the multiple MVMs. This peak statistics is another evidence supporting the existence of the multiple MVMs on the surface of ${\rm FeTe}_{x}{\rm Se}_{1-x}$. 

On the other hand, we examine the vortex cores with ZBPs as the second energy peaks in the experimental data \cite{2018arXiv181208995M}. The distribution of the second peaks in those vortices closest to zero energy is always empty near zero energy with $60\mu$eV gap. The second peaks can arise from the CdGM mode with the lowest non-zero energy~\cite{2019arXiv190102293K} or the hybridization of the MVMs. Fig.~\ref{ZBP_distribution}(g) demonstrates that the second peak stems only from the Majorana hybridization in the TB model simulation since the energies of the CdGM modes from the surface superconductivity are greater than $0.96$meV. However, it is difficult to distinguish the two different origins by merely looking at the energy peaks. Although it requires further examination, we suggest that the wide spreading of the second peaks in Fig.~\ref{ZBP_height_and_rate}(b) can result from many factors --- the coupling of the multiple CdGM modes nearby from the bulk superconductivity, the different strengths of the Majorana hybridization, and the inhomogeneous spatial distribution of the superconductor gap and chemical potential since the energy of the CdGM mode is roughly proportional to $\Delta^2/\mu$. 

	The emergence of the second peaks closest to zero energy can further explain the observation of the low ZBP rates~\cite{Wang2018Evidence} when the vortex cores in ${\rm FeTe}_{x}{\rm Se}_{1-x}$ were probed by lower-energy-resolution STM regardless the strength of the magnetic field. Consider one ZBP and another peak closest to zero energy. We note that a CdGM mode can lead to the second peak and breaks particle-hole symmetry of LDOS~\cite{PhysRevLett.80.2921}. The high-energy-resolution STM can clearly distinguish the two peaks as illustrated in Fig.~\ref{ZBP_height_and_rate}(d), whereas in the low-energy-resolution STM the two energy peaks merge to one peak away from zero energy as illustrated in Fig.~\ref{ZBP_height_and_rate}(e). In this regard, the ZBP cannot be observed by the low-energy-resolution STM when the energy of the second peak is roughly within the energy resolution. Thus, this is the reason that the low-energy-resolution STM has low ZBP rates. 

	

	 \vspace{0.5cm}



\noindent
{\bf ZBPs and vortex locations}	\\
Since the relation between the presence of the ZBPs and the Majorana hybridization on the surface is subtle, it has been studied~\cite{2018arXiv181208995M} that the presence of the ZBPs in the vortex cores does not exhibit simple correlations with spatial distributions of atoms, zero-magnetic-field in-gap bound states, and superconducting gaps. On the other hand, using time dependence measurements, we observe that the creeping vortex, which slightly moved on the surface of ${\rm FeTe}_{x}{\rm Se}_{1-x}$, has a significant change of the LDOS as shown in Fig.~\ref{creeping_vortex}.  Furthermore, our additional STM measurement shows that at $B=4T$ in some cases the LDOS of the vortex is not affected by other moving vortices when its surrounding vortex distribution is unchanged as shown in Fig.~\ref{experiment_4T_vortex_redistribution}. 
To understand how the LDOS in the vortices is affected by the Majorana hybridization physics, we study how a single moving vortex alters the LDOS in each vortex core in the field of view (FOV). We simulate the TB model of the iron-based superconductor in the high magnetic field (4T, $\bar{R}_v=24$nm). We start with the vortex distribution in Fig.~\ref{vortex_move}(a) and the vortex in the center of the FOV moves down $7$nm twice. The first movement (Fig.~\ref{vortex_move}(b)) dramatically changes the distribution of the ZBPs in the vortex cores and the ZBP rate $R_{\rm{ZBP}}$ from $47.6\%$ to $31.7\%$, while the second movement (Fig.~\ref{vortex_move}(c)) of the central vortex does not lead to the dramatic change of the ZBP rate. This significant change in the first movement demonstrates that the ZBP rate $R_{\rm{ZBP}}$ fluctuates in the wide range with the same average intervortex distance as shown in Fig.~\ref{ZBP_height_and_rate}(a).  

LDOS for each vortex core varies as the central vortex moves. The moving vortex, which initially possesses a ZBP (Fig.~\ref{vortex_move}(d)$\alpha$), does not have a ZBP after the movements. Surprisingly, this movement can also affect the LDOS of the vortex $120$nm away from the moving vortex (cf.~the coherence length $\xi=13.9$nm) as Fig.~\ref{vortex_move}(d)$\epsilon$ shows the ZBP appears after the first movement. It is known that the coupling between the moving MVM and the distant MVM is  exponentially suppressed; however, the surprising LDOS change stems from that the coupling between the distant MVM ($\delta$) and its nearby MVMs is \emph{slightly} altered by the phase change ($\omega_{ij}$ in Eq.\ref{phase coupling}) from the central vortex ($\alpha$) movement, since the distance between those vortices ($\alpha,\ \epsilon$) is less than the London penetration depth ($\sim500$nm). The vortices near the moving vortex have significant changes of the LDOS as shown in Fig.~\ref{vortex_move}(d)($\alpha,\beta,\gamma,\delta$), since in this short length scale the coupling of the moving MVM and its nearby MVMs are varied by the movement and the coupling change between the other MVMs stems from the phase change ($\omega_{ij}$) from the moving vortex. In the lower magnetic fields, the moving vortex affects the LDOS of its nearby vortices as shown in Fig.~\ref{distribution46_5nm},\ref{distribution33nm}. However, the LDOS of the vortices away from the moving vortex is not altered dramatically and there are slight changes in the ZBP rates. 

	Now we are back to our new experimental data of the moving vortices in Fig.~\ref{creeping_vortex},\ref{experiment_4T_vortex_redistribution}. First, as shown in Fig.~\ref{creeping_vortex} the LDOS of the creeping vortex exhibits a dramatic change after the creeping vortex moved $2$nm. This change can be explained by the change of the Majorana hybridization with the surrounding MVMs, because we have shown that the $7$nm vortex movement can significantly {\clr change} the LDOS in Fig.~\ref{vortex_move}(d)($\beta,\gamma$). Furthermore, since the spatial range of the FOV probed by STM for the creeping vortex in Fig.~\ref{creeping_vortex} is limited in $15$nm$\times 15$nm square, it is possible that the surrounding vortices, which were not observed by STM, also moved at the same time. The location change of the surrounding vortices can affect the LDOS of the creeping vortex. Second, as shown in Fig.~\ref{experiment_4T_vortex_redistribution} the LDOS of the observed vortex is unchanged after the movement of the vortices, which are away from the observed one. We have shown in Fig.~\ref{vortex_move}(d)($\epsilon,\zeta$) that the LDOS of the vortices away from the moving vortex have only slight changes. Since in the experimental data the smallest distance between the observed vortex and the moving vortices is around $50$nm, which is much longer than the Majorana coherence length, the LDOS of the observed vortex can be unchanged. Thus, the simulation of the vortex movement can qualitatively explain the observed LDOS in the vortices before and after the vortex movement.


\begin{figure}[t]
\begin{center}
\includegraphics[clip,width=0.99\columnwidth]{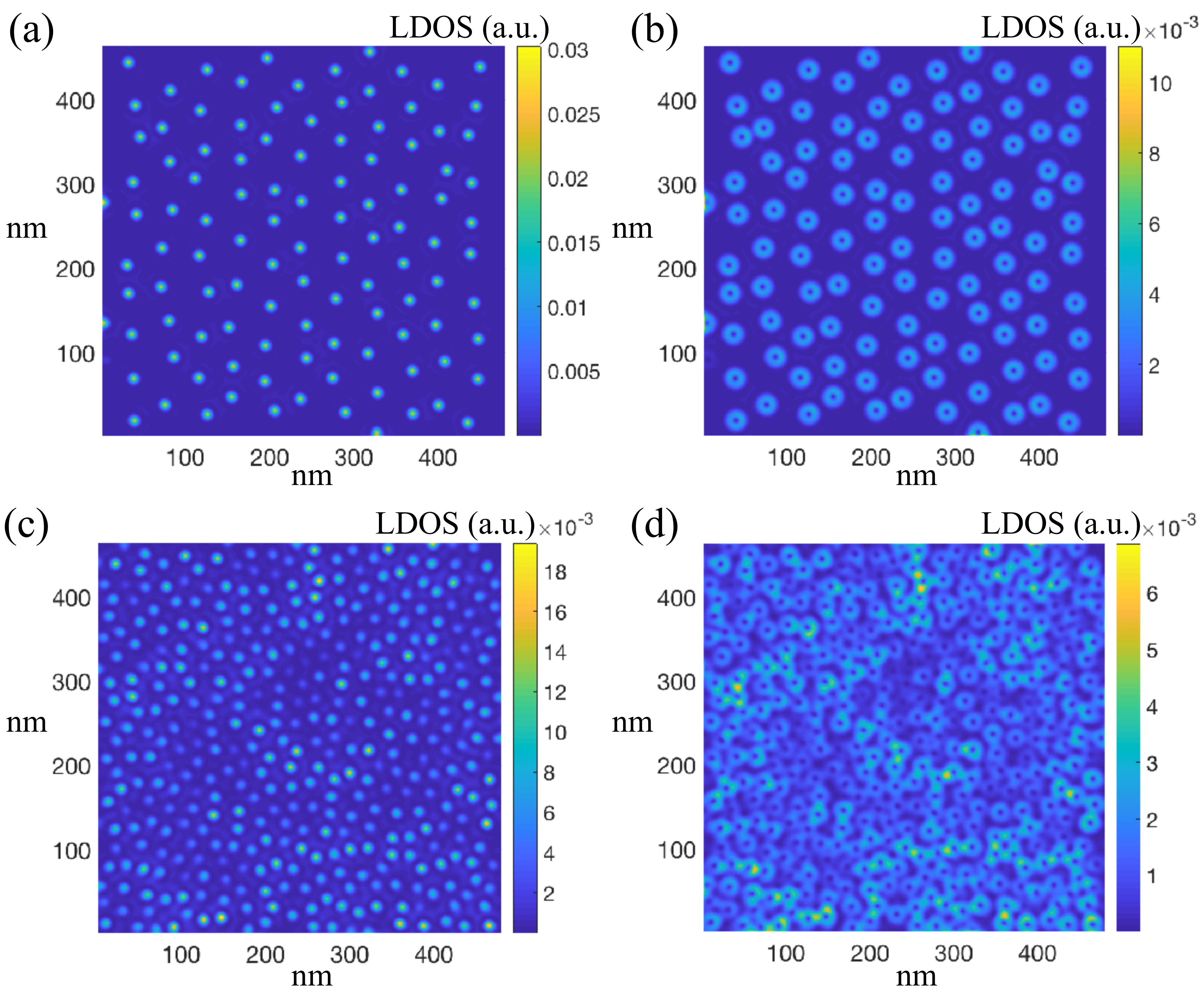}
\caption{  {\bf Spin-LDOS with energy within $\pm0.15$meV} (a,b) share the same FOV in Fig.~\ref{ZBP_distribution}(a) and show respectively the LDOS of spin up and down with the distorted vortex lattice distribution and the average intervortex distance $\bar{R}_v=46.5$nm. (c,d) share the same FOV in Fig.~\ref{ZBP_distribution}(e) and show respectively the LDOS of spin up and down with the glass vortex distribution and the average intervortex distance $\bar{R}_v=24$nm. The LDOS of spin up is similar with the LDOS including both spins in Fig.~\ref{ZBP_distribution}(a,e) due to the domination of spin up. The LDOS of spin down exhibits donut-shaped peaks encircling each vortex core as the signature of the multiple MVMs. }
\label{spin_distribution} 
\end{center}
\end{figure}



\vspace{0.5cm}


\noindent
{\bf Spin LDOS of multiple Majorana vortex modes} \\
We have shown the multiple clues of the recent STM measurements in agreement with the results of the TB model hosting hundreds of MVMs on the surface. However, the observation of the tunneling peaks in the vortices is not enough to conclude the Majorana nature on the surface of ${\rm FeTe}_{x}{\rm Se}_{1-x}$. It is crucial to predict and to observe the universal signature of the multiple MVMs even when the Majorana hybridization is strong and the ZBPs are removed.

 In the literature~\cite{PhysRevLett.115.177001} it has been proposed that the spin-polarized wavefunction of the isolated MVM oscillates spatially as the signature of the Majorana observation. On the one hand, practically the oscillation behavior might be difficult to be observed since the exponential decay of the Majorana wavefunction flattens the spin oscillation so the current STM might not have enough sensitivity to detect the oscillation away from the vortex cores. On the other hand, since the Majorana coherence length $\xi=13.9$nm is comparable with the average intervortex distance $\bar{R}_v$ from $19$nm to $46.5$nm, in the presence of the Majorana hybridization it is uncertain that the wavefunction (\ref{cont state}) of each individual MVM is kept in the original form. Thus, before experimentally looking for the spin signature of the multiple MVMs in the future, we first study the TB model of ${\rm FeTe}_{x}{\rm Se}_{1-x}$ to confirm the spin oscillation is not altered by the Majorana hybridization and check the requirement to observe the spin oscillation by the spin-selected STM probe. 
 
 	The spin oscillation of the isolated MVM wavefunction (\ref{cont state}) stems from the Bessel functions $J_0(k_F r)$ and $J_1(k_F r)$ for spin up and down respectively, where $r$ is the distance away from the vortex core. Therefore, due to the properties of the Bessel functions, the first peaks appear at $r=0$ and $r=1.84/k_F$ for spin up and down and the other oscillation peaks are exponentially suppressed. In the low magnetic field, the TB model shows similar spin spatial peaks for the vortices with and without ZBPs as shown in Fig.~\ref{spin_LDOS}. That is, for the isolated and hybridized MVMs, the LDOS of spin up exhibits peaks appearing at the vortex centers as well as the LDOS of spin down vanishes at the vortex centers and has peaks away from the centers for $r=1.84/k_F$. Although we focus on the spatial locations of the peaks, the energy peaks might be away from zero energy due to the Majorana hybridization. However, since as shown in Fig.~\ref{spin_LDOS} the peak height of spin down, which is suppressed by the exponential decay, is roughly 6 times smaller than the one of spin up, the spin down peak might be difficult to be detected by STM without high sensitivity. Alternatively, we integrate the LDOS within $\pm 0.15$meV energy range, which is greater than the current STM energy resolution. Fig.~\ref{spin_distribution} shows the FOV for the two different spin distributions in the weak ($1$T:(a,b)) and strong ($4$T:(c,d)) magnetic fields. It is similar with the conventional STM probe that the spin up peaks in (a,c) indicate the locations of the vortex centers. Differently, as shown in Fig.~\ref{spin_distribution}(b,d) the FOV of spin down shows each vortex core is encircled by a donut-shaped peak since the LDOS spatial peaks of spin down appear away from the vortex cores. In the high magnetic field, due to the high density of the vortices, the donuts are connected to other neighbors. Although the MVMs strongly hybridize, preserving the MVM wavefunction leads to the observation of the donuts. This observation will be a smoking gun to show the ${\rm FeTe}_{x}{\rm Se}_{1-x}$ surface possessing the multiple MVMs. 



\vspace{0.5cm}

\noindent
\textbf{\large Discussion}\\
The decreasing fraction of vortices possessing ZBPs on the surface of ${\rm FeTe}_{x}{\rm Se}_{1-x}$ can be successfully explained by the hybridization of the MVMs with disordered vortex distributions. In other words, the recent observation of the decreasing ZBP rate supports the existence of multiple MVMs. While the oscillation of the Majorana coupling as a function of the intervortex distance has been erased by the distorted triangular lattice and glass of the vortex distributions, the Majorana coupling of the exponential decay as a function of the intervortex distance survives. Since the Majorana coupling increases with the increasing magnetic field, the stronger hybridization moves more MVMs away from zero energy and then reduces the ZBP rate. In other words, the reducing ZBP rate is the first evidence showing the hybridizations of multiple MVMs. Furthermore, by analyzing the energy peak statistics from the STM measurement we find the numbers of the first LDOS peaks closest to zero energy, in the vortex cores not possessing ZBPs, has a pyramid-shaped distribution except for zero energy. This distribution, which is consistent with the TB simulation, can link to the Majorana hybridization moving the peaks away from zero energy.




	The interplay between the LDOS and the detailed couplings of the MVMs is fascinatingly complicated. The slight movement ($\sim 14$nm coherence length) of the single vortex can significantly affect the LDOS of the vortex cores near the moving vortex, while in the high magnetic field the vortex movement can alter the absence or presence of some ZBPs in vortex cores away from the moving vortex (distance less than the London penetration depth). It has been shown that manipulating the Majorana movement can manifest quantum logic gates~\cite{2019arXiv190106083L,Alicea_Majorana_wire} for quantum computing. Encouraged by the recent success controlling the movement of a single superconducting vortex~\cite{Ge:2016aa}, we believe that ${\rm FeTe}_{x}{\rm Se}_{1-x}$ can be an ideal platform to achieve Majorana quantum computing. 
	

	Since the spin textures of the MVMs survive in the presence of the Majorana hybridization, this universal spin feature of multiple hybridized MVMs probed by the spin-selected STM will be a smoking gun of scalable Majorana qubits in the iron-based superconductor platform. We predict that the multiple MVMs exhibit the spatial donut-shaped distribution of spin down surrounding the vortex cores. After the experimental confirmation, to have scalable Majorana qubits, we apply low magnetic field in ${\rm FeTe}_{x}{\rm Se}_{1-x}$ so that each MVM is well separated to stay at zero energy and the quantum information stored in the Majorana qubits can be safely protected by topology. Having a platform of the scalable Majorana qubits fulfills one key of the DiVincenzo criteria for quantum computing. This platform can potentially become a primary step to build scalable quantum gates formed by MVMs and to read out Majorana qubits in the future. \\

\vspace{0.5cm}

\noindent
\textbf{\large Methods}\\
	The Supplementary materials provides additional results from STM measurements as well as a detailed and systematic three-dimensional TB model construction describing the physics of multiple MVMs on the ${\rm FeTe}_{x}{\rm Se}_{1-x}$ surface. On the one hand, the experimental methodology was already written in the recent work~\cite{2018arXiv181208995M}. On the other hand, as our numerical analysis is lengthy, here we point out the keys of the approach. 
	
	We distill the main physical features of ${\rm FeTe}_{x}{\rm Se}_{1-x}$ to capture the Majorana nature by only considering the s-wave pairing surface Dirac cone, since on the superconducting surface Dirac cone the two characteristic lengths, the Majorana coherence length $\xi=\nu_F/\Delta_0$ and the Fermi length $1/k_F$, determine the main Majorana physics. First, by adopting the physical values of the SC gap $\Delta_0=1.8$meV, the Fermi velocity of the Dirac cone $\nu_F=25$nm$\cdot$meV, $k_F=0.2$nm$^{-1}$, its chemical potential $\mu=\nu_F/k_F=5$meV, we build a TB model describing a single MVM on the (001) surface. When the simulation is extended to include hundreds of MVMs on the same surface, the magnetic fluxes through the vortices significantly affect the couplings of the MVMs. We introduce the vector potential in the TB model describing $h/2e$ magnetic flux through each vortex with the London penetration depth $\lambda=500$nm. By performing the Lanczos algorithm, we find all of the energy states within $\pm 0.25$meV energy range. Using these energy states, we compute the LDOS in the vortex cores with effective temperature $T=20\mu$eV, which leads to the FWHM close to the experimental observation. The criterion to determine ZBPs in the vortex cores is that the energy peak location of the LDOS is within $\pm 20\mu$eV energy range. 
	
%
%
%
%
%
%

\vspace{0.5cm}
\noindent
\textbf{Acknowledgments}: The authors thank S. Das Sarma, M. Franz, L. Kong, D. Liu, T. Liu, X. Liu, P. A. Lee, T. Posske, D. Wang, H.-H. Wen, R. Wiesendanger, and H. Zheng for the helpful discussions and comments. 
\textbf{Funding}: C.-K.C. and F.-C.Z. are supported by the Strategic Priority Research Program of the Chinese Academy of Sciences (Grant XDB28000000). T.M. and T.H. are supported by CREST project JPMJCR16F2 from Japan Science and Technology Agency. T.M. is also supported by a Grants-in-Aid for Scientific Research (KAKENHI) (numbers 19H01843) by the Japan Society for the Promotion of Science (JSPS).  F.-C.Z. is also supported by Beijing Municipal Science $\&$ Technology Commission (No. Z181100004218001), National Science Foundation of China (Grant No.11674278), National Basic Research Program of China (No.2014CB921203). \textbf{Author contributions}: C.-K.C and Y.H. built and performed the tight-binding model. T.M. and T.H. analyzed the experimental data. C.-K.C. wrote the manuscript. All authors discussed the results and contributed to the manuscript. \textbf{Competing interests}: The authors declare that they have no competing interests. \textbf{Data and materials availability}: All data needed to evaluate the conclusions in the paper are present in the paper and/or the Supplementary Materials. Additional data related to this paper may be requested from the authors. 

\vspace{0.5cm}
\noindent
\textbf{\large Supplementary materials}\\
Appendix A: 3D tight-binding model of FeTe$_x$Se$_{1-x}$\\
Appendix B: Simulation details \\
Appendix C: Particle-hole symmetry broken by Majorana hybridization\\
Appendix D: Simulation for a moving vortex \\
Appendix E: STM measurement of creeping vortex and vortex relocation \\
Appendix F: Simulation for LDOS line profiles of ZBP and non-ZBP vortices \\
Appendix G: Inhomogeneous superconducting gaps and chemical potentials \\
Appendix H: ZBP heights \\




 

%
%
%
%
%
%
%
%
%
%



\appendix

\setcounter{figure}{0}
\makeatletter 
\renewcommand{\thefigure}{S\@arabic\c@figure} 

\makeatother

\bibliography{BibMajorana,TOPO_v14}

\begin{thebibliography}{10}

\bibitem{Kitaev2001}
A.~{Kitaev}, {\it Physics Uspekhi\/} {\bf 44}, 131 (2001).

\bibitem{RMP_braiding}
C.~Nayak, S.~H. Simon, A.~Stern, M.~Freedman, S.~Das~Sarma, {\it Rev. Mod.
  Phys.\/} {\bf 80}, 1083 (2008).

\bibitem{Gil_Majorana_wire}
Y.~Oreg, G.~Refael, F.~von Oppen, {\it Phys. Rev. Lett.\/} {\bf 105}, 177002
  (2010).

\bibitem{Sau_semiconductor_heterostructures}
J.~D. Sau, R.~M. Lutchyn, S.~Tewari, S.~Das~Sarma, {\it Phys. Rev. Lett.\/}
  {\bf 104}, 040502 (2010).

\bibitem{Roman_SC_semi}
R.~M. Lutchyn, J.~D. Sau, S.~Das~Sarma, {\it Phys. Rev. Lett.\/} {\bf 105},
  077001 (2010).

\bibitem{Fu2008Superconducting}
L.~Fu, C.~L. Kane, {\it Phys. Rev. Lett.\/} {\bf 100}, 096407 (2008).

\bibitem{Zhang:2018aa}
H.~Zhang, {\it et~al.\/}, {\it Nature\/} {\bf 556}, 74 EP  (2018).

\bibitem{Wang2018Evidence}
D.~Wang, {\it et~al.\/}, {\it Science\/} {\bf 362}, 333 (2018).

\bibitem{2018arXiv181208995M}
T.~Machida, {\it et~al.\/}, {\it Nature Materials\/} {\bf 18}, 811 (2019).

\bibitem{PhysRevX.8.041056}
Q.~Liu, {\it et~al.\/}, {\it Phys. Rev. X\/} {\bf 8}, 041056 (2018).

\bibitem{DiVincenzo:2000aa}
D.~P. DiVincenzo, {\it Fortschritte der Physik\/} {\bf 48}, 771 (2000).

\bibitem{Chiu_strong_inter_Majorana}
C.-K. Chiu, D.~I. Pikulin, M.~Franz, {\it Phys. Rev. B\/} {\bf 91}, 165402
  (2015).

\bibitem{Xu2015Experimental}
J.-P. Xu, {\it et~al.\/}, {\it Phys. Rev. Lett.\/} {\bf 114}, 017001 (2015).

\bibitem{Sun2016Majorana}
H.-H. Sun, {\it et~al.\/}, {\it Phys. Rev. Lett.\/} {\bf 116}, 257003 (2016).

\bibitem{Law2009Majorana}
K.~T. Law, P.~A. Lee, T.~K. Ng, {\it Phys. Rev. Lett.\/} {\bf 103}, 237001
  (2009).

\bibitem{CAROLI1964307}
C.~Caroli, P.~D. Gennes, J.~Matricon, {\it Physics Letters\/} {\bf 9}, 307
  (1964).

\bibitem{10.1093/nsr/nwy142}
N.~Hao, J.~Hu, {\it National Science Review\/} {\bf 6}, 213 (2018).

\bibitem{Zhang2018Observation}
P.~Zhang, {\it et~al.\/}, {\it Science\/} {\bf 360}, 182 (2018).

\bibitem{Chen:2018aa}
M.~Chen, {\it et~al.\/}, {\it Nature Communications\/} {\bf 9}, 970 (2018).

\bibitem{2019arXiv190406124Z}
S.~{Zhu}, {\it et~al.\/}, {\it arXiv e-prints arXiv:1904.06124\/}  (2019).

\bibitem{Chen_2019}
C.~Chen, {\it et~al.\/}, {\it Chinese Physics Letters\/} {\bf 36}, 057403
  (2019).

\bibitem{doi:10.1063/1.5049619}
T.~Machida, Y.~Kohsaka, T.~Hanaguri, {\it Review of Scientific Instruments\/}
  {\bf 89}, 093707 (2018).

\bibitem{2019arXiv190901686C}
X.~{Chen}, {\it et~al.\/}, {\it arXiv e-prints\/} p. arXiv:1909.01686 (2019).

\bibitem{PhysRevLett.103.107001}
M.~Cheng, R.~M. Lutchyn, V.~Galitski, S.~Das~Sarma, {\it Phys. Rev. Lett.\/}
  {\bf 103}, 107001 (2009).

\bibitem{:2010tb}
M.~Cheng, R.~M. Lutchyn, V.~Galitski, S.~Das~Sarma, {\it Phys. Rev. B\/} {\bf
  82}, 094504 (2010).

\bibitem{Liu_2015}
T.~Liu, M.~Franz, {\it Physical Review B\/} {\bf 92}, 134519 (2015).

\bibitem{PhysRevB.95.224505}
S.~Kolenda, C.~S\"urgers, G.~Fischer, D.~Beckmann, {\it Phys. Rev. B\/} {\bf
  95}, 224505 (2017).

\bibitem{PhysRevB.81.180503}
H.~Kim, {\it et~al.\/}, {\it Phys. Rev. B\/} {\bf 81}, 180503 (2010).

\bibitem{PhysRevLett.115.177001}
T.~Kawakami, X.~Hu, {\it Phys. Rev. Lett.\/} {\bf 115}, 177001 (2015).

\bibitem{PhysRevB.94.224501}
L.-H. Hu, C.~Li, D.-H. Xu, Y.~Zhou, F.-C. Zhang, {\it Phys. Rev. B\/} {\bf 94},
  224501 (2016).

\bibitem{PhysRevLett.111.136401}
R.~R. Biswas, {\it Phys. Rev. Lett.\/} {\bf 111}, 136401 (2013).

\bibitem{PhysRevB.82.184506}
T.~Klein, {\it et~al.\/}, {\it Phys. Rev. B\/} {\bf 82}, 184506 (2010).

\bibitem{SERRIERGARCIA2017109}
L.~Serrier-Garcia, M.~Timmermans, J.~V. de~Vondel, V.~V. Moshchalkov, {\it
  Physica C: Superconductivity and its Applications\/} {\bf 533}, 109  (2017).
  Ninth international conference on Vortex Matter in nanostructured
  Superdonductors.

\bibitem{RAMAN:1927aa}
C.~V. RAMAN, C.~M. SOGANI, {\it Nature\/} {\bf 119}, 601 (1927).

\bibitem{PhysRevB.81.184511}
S.~Khim, {\it et~al.\/}, {\it Phys. Rev. B\/} {\bf 81}, 184511 (2010).

\bibitem{PhysRevB.98.144519}
C.~Berthod, {\it Phys. Rev. B\/} {\bf 98}, 144519 (2018).

\bibitem{2019arXiv190102293K}
L.~{Kong}, {\it et~al.\/}, {\it arXiv e-prints arXiv:1901.02293\/}  (2019).

\bibitem{PhysRevLett.80.2921}
N.~Hayashi, T.~Isoshima, M.~Ichioka, K.~Machida, {\it Phys. Rev. Lett.\/} {\bf
  80}, 2921 (1998).

\bibitem{2019arXiv190106083L}
C.-X. {Liu}, D.~E. {Liu}, F.-C. {Zhang}, C.-K. {Chiu}, {\it arXiv e-prints\/}
  p. arXiv:1901.06083 (2019).

\bibitem{Alicea_Majorana_wire}
J.~Alicea, Y.~Oreg, G.~Refael, F.~von Oppen, M.~P.~A. Fisher, {\it Nature
  Physics\/} {\bf 7}, 412 (2011).

\bibitem{Ge:2016aa}
J.-Y. Ge, {\it et~al.\/}, {\it Nature Communications\/} {\bf 7}, 13880 EP
  (2016).

\bibitem{Zhang:2009aa}
H.~Zhang, {\it et~al.\/}, {\it Nat Phys\/} {\bf 5}, 438 (2009).

\bibitem{PhysRevB.79.054503}
S.~Li, {\it et~al.\/}, {\it Phys. Rev. B\/} {\bf 79}, 054503 (2009).

\bibitem{deGennesbook}
P.~G. de~Gennes, {\it Superconductivity Of Metals And Alloys\/} (Westview
  Press, 1999).

\bibitem{Chiu:2011fk}
C.-K. Chiu, M.~J. Gilbert, T.~L. Hughes, {\it Phys. Rev. B\/} {\bf 84}, 144507
  (2011).

\bibitem{PhysRevB.86.155146}
D.~J.~J. Marchand, M.~Franz, {\it Phys. Rev. B\/} {\bf 86}, 155146 (2012).

\end{thebibliography}

\bibliographystyle{Science}

 \clearpage
\newpage

\begin{center}
\textbf{
\large{Supplementary Materials for}}
\vspace{0.4cm} 

\textbf{
\large{
``Scalable Majorana vortex modes in iron-based superconductors" } 
}
\end{center}

\vspace{0.1cm}

\begin{center}
\textbf{Authors:} 
Ching-Kai Chiu,
T. Machida, Yingyi Huang, T. Hanaguri, Fu-Chun Zhang
\end{center}

\vspace{0.5cm}

In these supplementary materials we present the details of the vortex tight-binding model for the iron-based superconductors ${\rm FeTe}_{x}{\rm Se}_{1-x}$.

\section{3D tight-binding model of ${\rm FeTe}_{x}{\rm Se}_{1-x}$}  \label{Simulation} 


Many physical mechanisms are present on the surface of ${\rm FeTe}_{x}{\rm Se}_{1-x}$ probed by the recent experiments~\cite{Zhang2018Observation,Wang2018Evidence,2018arXiv181208995M}. Since the focus is on the quantum states with energy within $\pm 0.2$meV, only Majorana physics emerges in this energy range. The reason is that the superconductor gap of the bulk and surface is much greater than $0.2$meV~\cite{Wang2018Evidence} and the low energy modes (CdGM modes) located in the vortex cores possess energy higher than $0.2$meV~\cite{2019arXiv190102293K}. Hence, the low-energy physics already excludes the CdGM modes and the quasiparticles with energy higher than the superconductor gaps. Only MVMs emerge in the low-energy physics.   
The non-trivial topological superconductivity of ${\rm FeTe}_{x}{\rm Se}_{1-x}$, which hosts MVMs, stems from the surface Dirac cone. Eq.~\ref{cont model} in the manuscript shows that the phase winding of the vortex aligns the spin rotation texture of the surface Dirac cone so that the vortex can host an MVM. Although the $d-$wave orbitals are involved in the basis of the surface Dirac cone~\cite{10.1093/nsr/nwy142}, the emerging of the MVMs results from the spin texture of the Dirac cone, regardless of the choice of the orbitals. To capture the salient physics of multiple MVMs here, we start with the model of a non-trivial 3D topological insulator (TI) hosting a surface Dirac cone~\cite{Zhang:2009aa}, which is identical to the one in ${\rm FeTe}_{x}{\rm Se}_{1-x}$. 
The TI Hamiltonian in momentum space is written as  
\begin{align}
H_{\rm{TI}}=&\nu_FM(\bk) \rho_z \sigma_0+\nu_F \sin k_x a \rho_x \sigma_x + \nu_F \sin k_y a \rho_x \sigma_y\nonumber \\
& +\nu_F \sin k_z  a\rho_x\sigma_z -\mu \rho_0 \sigma_0,
\end{align}
where $M(\bk)=m-\cos k_x a- \cos k_y a -\cos k_z a$, $\nu_F=25$nm$\cdot$meV, $k_F=0.2$nm$^{-1}$, and chemical potential $\mu=\nu_F/k_F=5$meV based on the experimental measurement~\cite{Zhang2018Observation,Wang2018Evidence}. The Pauli matrices $\sigma_\alpha$ and $\rho_\beta$ represent spin and orbital degrees of freedom respectively. 
Although for ${\rm FeTe}_{x}{\rm Se}_{1-x}$ its space group is $P4/nmm$, and the lattice constants have been known $a=b=3.8\AA$ $c=6.1\AA$~\cite{PhysRevB.79.054503}, we use cubic lattice and choose the lattice constant $a=2$nm in all of the directions to simplify the problem and focus on the physics of the surface Dirac cone, since the value of the lattice constant does not affect the low-energy Majorana physics.  

By introducing s-wave superconductivity, the TI Hamiltonian is extended to the BdG Hamiltonian
\begin{align}
H_{\rm{BdG}}=&\nu_FM(\bk) \tau_z \rho_z \sigma_0+\nu_F \sin k_x a \tau_0 \rho_x \sigma_x+ \nu_F \sin k_y a \tau_z \rho_x \sigma_y
\nonumber \\ 
& +\nu_F \sin k_z a \tau_0\rho_x\sigma_z -\mu \tau_z\rho_0 \sigma_0+\Delta_0 \tau_y \rho_0 \sigma_y,
\end{align}
where the value of the s-wave surface superconducting gap ($\Delta_0\approx 1.8$meV) has been measured in the experiments~\cite{Zhang2018Observation,Wang2018Evidence,2018arXiv181208995M,Chen:2018aa}. Now we have the estimated values of the two Majorana characteristic lengths --- Majorana coherence length $\xi=\nu_F/\Delta=13.9$nm and the Fermi wavelength $1/k_F=\nu_F/\mu=5$nm. Hence, the Fermi velocity $\nu_F$ of the surface Dirac cone is an essential quantity to determine the two characteristic lengths of the MVM. The primary platform we consider is the $(001)$ surface with vortices without translation symmetries; in this regard, the Bogoliubov-de Gennes (BdG) Hamiltonian can be rewritten in real space 
\begin{align}
\hat{H}_{\rm{BdG}}&= \sum_{\br} 
\bma 
C^\dagger_\br & C_{\br}
\ema
\bma
H_0(m) & -i\Delta_0 \tau_0 \sigma_y \\
i\Delta_0 \tau_0 \sigma_y & -H_0^*(m)
\ema
\bma 
C_{\br} \\
C^\dagger_{\br}
\ema  \nonumber \\
&+\sum_{\br,\bm{\delta}} 
\bma 
C^\dagger_\br & C_\br
\ema
\bma
H_{\rm{nn}}(\bm{\delta}) & 0 \\
0 & -H_{\rm{nn}}^*(\bm{\delta})
\ema
\bma 
C_{\br+\bm{\delta}} \\
C^\dagger_{\br+\bm{\delta}}
\ema,
\end{align}
where the on-site part $H_0(m)=\nu_F m \rho_z \sigma_0 - \mu  \rho_o \sigma_0 $, $\bm{\delta}=\pm a \hat{x},\pm a \hat{y},\pm a \hat{z}$ indicates the three directions of the hopping, and the nearest neighbor hopping part $H_{\rm{nn}}(\pm a \hat{n})=-\nu_F \rho_z \sigma_0/2 \pm i \nu_F \rho_x \sigma_n/2$ for $n=x,y,z$. To reduce the number of the layers in the $z$ direction for the numerical simulation, we specifically choose $m=2$ so that the surface Dirac cones with small momentum can be almost located on the top and bottom layers.  On these layers, the neutral point of the Dirac cone is located at $(0,0)$ in the $(001)$ surface BZ, which is consistent with ${\rm FeTe}_{x}{\rm Se}_{1-x}$.  

\begin{widetext} 
Since ${\rm FeTe}_{x}{\rm Se}_{1-x}$ is a type-II superconductor, multiple Abrikosov vortices can be present with magnetic flux through the superconductor. In the presence of the vortices, the BdG Hamiltonian is in the form of 
\begin{align}
\hat{H}_{\rm{BdG}}&= \sum_{\br} 
\bma 
C^\dagger_\br & C_{\br}
\ema
\bma
H_0(m) & -i\Delta(\br) \tau_0 \sigma_y \\
i\Delta^*(\br) \tau_0 \sigma_y & -H_0^*(m)
\ema
\bma 
C_{\br} \\
C^\dagger_{\br}
\ema  \nonumber \\
&+\sum_{\br,\bm{\delta}} 
\bma 
C^\dagger_\br & C_\br
\ema
\bma
H_{\rm{nn}}(\bm{\delta})e^{-\frac{ie}{\hbar}\int_{\br}^{\br+\bm{\delta}} \bm{A}(\br)\cdot d\bm{l}} & 0 \\
0 & -H^*_{\rm{nn}}(\bm{\delta})e^{\frac{ie}{\hbar}\int_{\br}^{\br+\bm{\delta}} \bm{A}(\br)\cdot d\bm{l}}
\ema
\bma 
C_{\br+\bm{\delta}} \\
C^\dagger_{\br+\bm{\delta}}
\ema, \label{main BdG Hamiltonian}
\end{align}
where the superconductor gap $\Delta(\br)$ has an additional phase stemming from the vortices and the nearest neighbor hopping is modified by the Peierls substitution integral with the vector potential $\bm{A}(\br)$ describing the magnetic flux. In order to perform the numerical simulation, it is necessary to know the explicit expressions of $\Delta(\br)$ and $\bm{A}(\br)$. 
\end{widetext}
First, we consider a single vortex located at $\br=(x,y)=(0,0)$ along the $z$ direction and then the superconducting gap is given by  
\bee
\Delta( \bm{r})=\Delta_0 \tanh(r/\xi_0)e^{i\theta},
\ee
where $\xi_0$ is the superconducting coherence length. According to the experimental results (refer to Fig.~2(g,h) in \cite{2018arXiv181208995M}), the superconducting gap appears roughly $5$nm away from the vortex cores, so $\xi_0=4.6$nm is chosen for the simulation. We note that the Majorana coherence length $\xi=13.9$nm, which is the length scale of the MVM, and the superconducting coherence length $\xi_0$ represent the different physical entities. To obtain the vector potential $\bm{A}(\br)$, we start with the magnetic field in the $z$ direction, which concentrates in the vortex core and can be described by the London equation~\cite{deGennesbook}
\bee
B_z(\br)-\lambda^2 \nabla^2 B_z(\br)= \Phi_0 \delta (\br), 
\ee
where $\Phi_0=h/2e$ indicates the $\pi$-flux of the total magnetic flux on the entire surface and $\lambda$ is the London penetration depth. For ${\rm FeTe}_{x}{\rm Se}_{1-x}$, the London penetration length $\lambda\approx 500$nm~\cite{PhysRevB.95.224505,PhysRevB.81.180503}. Since the London penetration length is much greater than the two characteristic lengths of the MVM ($\xi=13.9$nm and $1/k_F=5$nm), the magnetic field is diluted in the characteristic length scale. It is reasonable to neglect the magnetic flux effect for a single MVM in the continuum model in the manuscript. However, the magnetic flux plays a pivotal role in the physics of multiple MVMs; we still include the magnetic flux effect in the TB model with a single vortex.  

\begin{figure}[t]
\begin{center}
\includegraphics[clip,width=0.99\columnwidth]{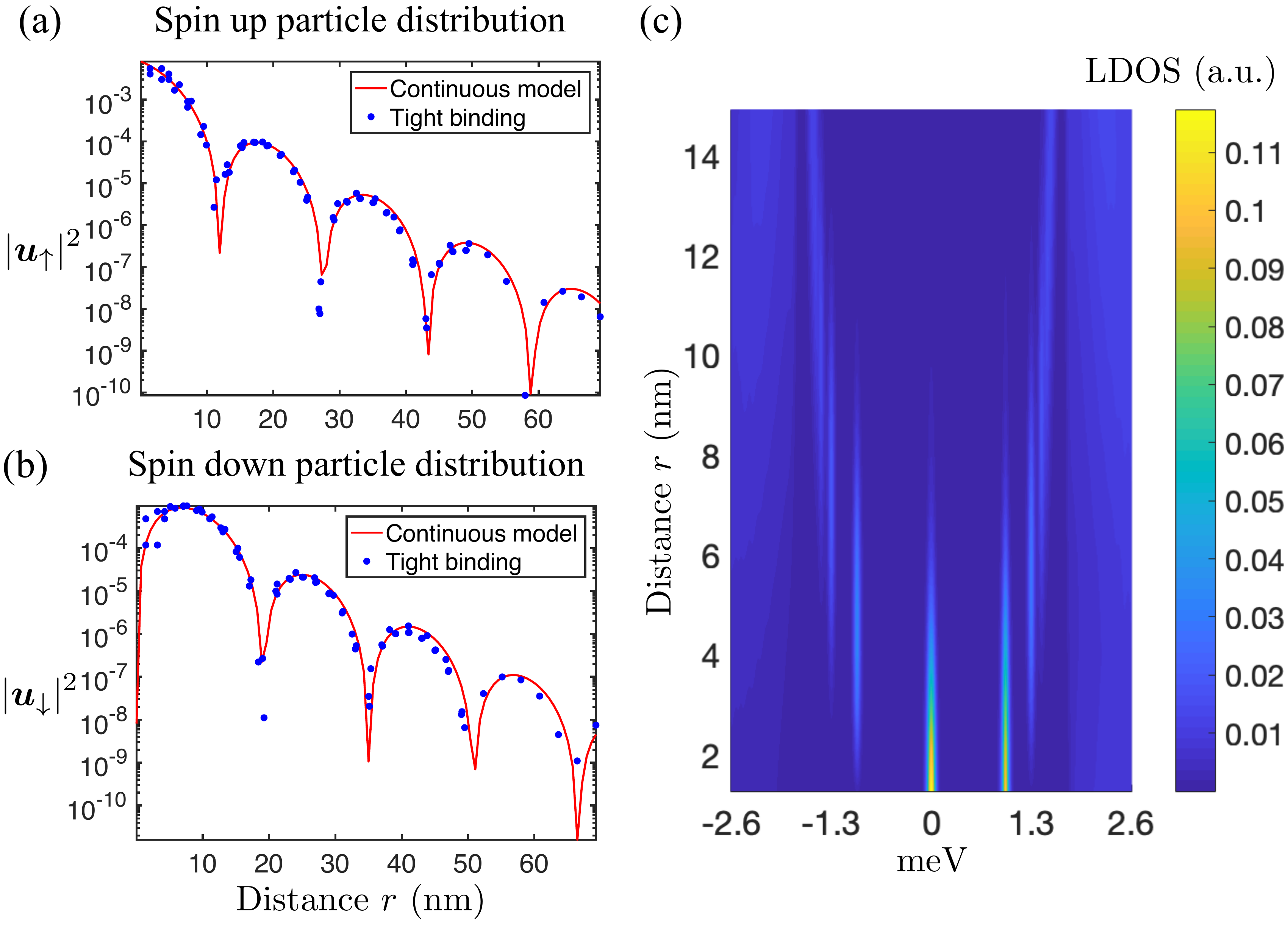}
\caption{ {\bf The line profile of the Majorana wavefunction} The spin up (a) and down (b) distributions of the particle part show that the continuum model (\ref{cont model}) and the tight binding model (\ref{main BdG Hamiltonian}) are in great agreement. In the continuum model, by eq.~\ref{cont state}, $|u_\uparrow|^2=e^{-2r/\xi}J_0^2(k_F r)$ and $|u_\downarrow|^2=e^{-2r/\xi}J_1^2(k_F r)$. For the tight binding model, we numerically find the MVM trapped in the vortex  located at the center of the $120\times144\times10$ (nm) system along the $z$ direction. (c) The line profile in the wide energy range shows the presence of the superconductor gap and the CdGM modes away from zero energy. 
 }
\label{one_vortex_up_down} 
\end{center}
\end{figure}

By solving the London equation, the magnetic field in the $z$ direction is given by  
\bee 
B_z(r)=\frac{\Phi_0}{2\pi \lambda^2}K_0(r/\lambda),
\ee
where $K_i(r/\lambda)$ is the modified Bessel function of the second kind. Furthermore, the total magnetic flux through the disk with the center at $(0,0)$ and the radius $r$ has the expression of 
\bee
\phi(r)=2\pi \int_0^r B_z(r')r'dr' =\big ( 1-\frac{r}{\lambda}K_1(r/\lambda) \big ) \Phi_0.
\ee
As $r\rightarrow \infty$, the asymptotic behavior of the modified Bessel function leads to $\phi\rightarrow \Phi_0$ indicating the $\pi$-flux forming the single Abrikosov vortex. By using the relation $\phi(r)=\oint \bm{A}(\bm{r})\cdot d\bm{l}$ in the closed integral path with the fixed radius $r$, the vector potential can be explicitly written as  
\bee
\bm{A}(\br)=\Phi_0(\frac{y}{r^2}\hat{x}-\frac{x}{r^2}\hat{y})(1-\frac{r}{\lambda}K_1\big (\frac{r}{\lambda})\big)
\ee
Now all of the parameters and functions in the BdG Hamiltonian (\ref{main BdG Hamiltonian}) with a single vortex are given; the wavefunction of the MVM trapped in the vortex on the surface of  ${\rm FeTe}_{x}{\rm Se}_{1-x}$ can be solved in this TB model. Fig.~\ref{one_vortex_up_down}(a,b) shows the TB model  (\ref{main BdG Hamiltonian}) and the continuum model (\ref{cont model}) are in great agreement although the superconducting gap in the continuum model, which does not vanish near the vortex core (no $ \tanh(r/\xi_0)$ dependent), is slightly different from the superconducting gap in the tight binding model. We note that the MZM wavefunctions are almost identical in the presence and absence of the magnetic flux  since the spatial range $r$ is much less than the London penetration depth~\cite{Chiu:2011fk}. Furthermore, the first three lowest energies of the CdGM modes trapped in the vortex in the TB are given by $\pm 0.96,\ \pm1.30,\ \pm1.43$meV as shown in Fig.~S1(c). The energy levels of the CdGM modes are in order of $\Delta^2/\mu$, it is not necessary to have equal energy spacing, when the CdGM energy is close to the superconductor gap. 



Since the surface of the type-II superconductor practically hosts more than one Abrikosov vortex, the physics of a single vortex is not enough to capture the mechanism of the topological superconductivity on the entire surface. Therefore, we consider multiple Abrikosov vortices located at $\br_j=(x_j,y_j)$ and then the superconducting gap is related to the locations of the vortices  
\bee
\Delta(\br)=\Delta_0\prod_j \tanh(|\br-\br_j|/\xi_0)\frac{x-x_j+(y-y_j)i}{|\br-\br_j|} \label{Delta}
\ee
The vector potential for the multiple vortices is given by 
\bee
\bm{A}(\br)=\Phi_0\sum_{j}\frac{(y-y_j)\hat{x}+(x-x_j)\hat{y}}{|\br-\br_j|}\Big(\frac{1}{|\br-\br_j|}-\frac{1}{\lambda}K_1(\frac{|\br-\br_j|}{\lambda})\Big) \label{vector potential all}
\ee
Thus, the complete BdG Hamiltonian of the TB model for multiple vortices, which are given, can be solved in principle.

\section{Simulation details} \label{simulation details}

\begin{figure}[t!]
\begin{center}
\includegraphics[clip,width=0.99\columnwidth]{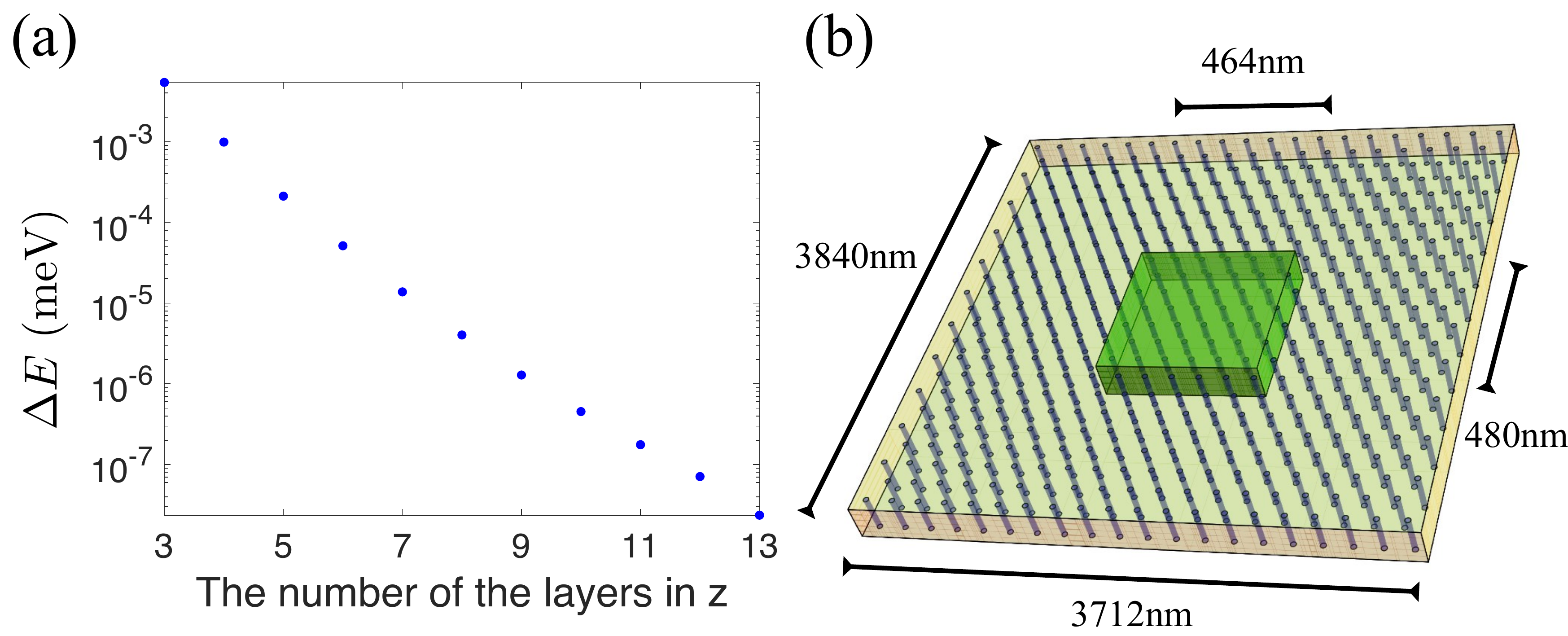}
\caption{ {\bf The system size of the 3D TB model for the simulation} (a) shows the hybridization energy of the two MVMs linked by the vortex on the top and bottom layers exhibits exponential decay as the number of the layers in the $z$ direction $L_z$ increases. (b) the green box represents the real size system physically including the vortices (blue) inside, while the yellow box represents the virtual size system contributing the phase effect from the vortices inside. 
 }
\label{simulation_size} 
\end{center}
\end{figure}

The major problem for the simulation is the choice of the system size ($L_x,L_y,L_z$) in the unit of the lattice constant. We expect to have $L_z$ as small as possible without losing the Majorana physics so that the size of the (001) surface can be extended to include more vortices. The two MVMs connected through the vortex along the $z$ direction can hybridize. This hybridization is expected to be strongly suppressed so the ZBP for an MVM can be seen on the surface. Here, we choose $m=2$ in the TB model (\ref{main BdG Hamiltonian}) so that the surface Dirac cone has a rapidly exponential decay along the $z$ direction~\cite{PhysRevB.86.155146}. Fig.~\ref{simulation_size}(a) shows the energy hybridization $\Delta E$ of the only two MVMs on the top and bottom layers decays exponentially as a function of $L_z$. Since for $L_z=3$ the energy hybridization $\Delta E=3\mu$eV is within $\pm 20\mu$eV energy range of the ZBP criterion, we safely choose $L_z=3$ for the simulation.

\begin{figure}[t!]
\begin{center}
\includegraphics[clip,width=0.99\columnwidth]{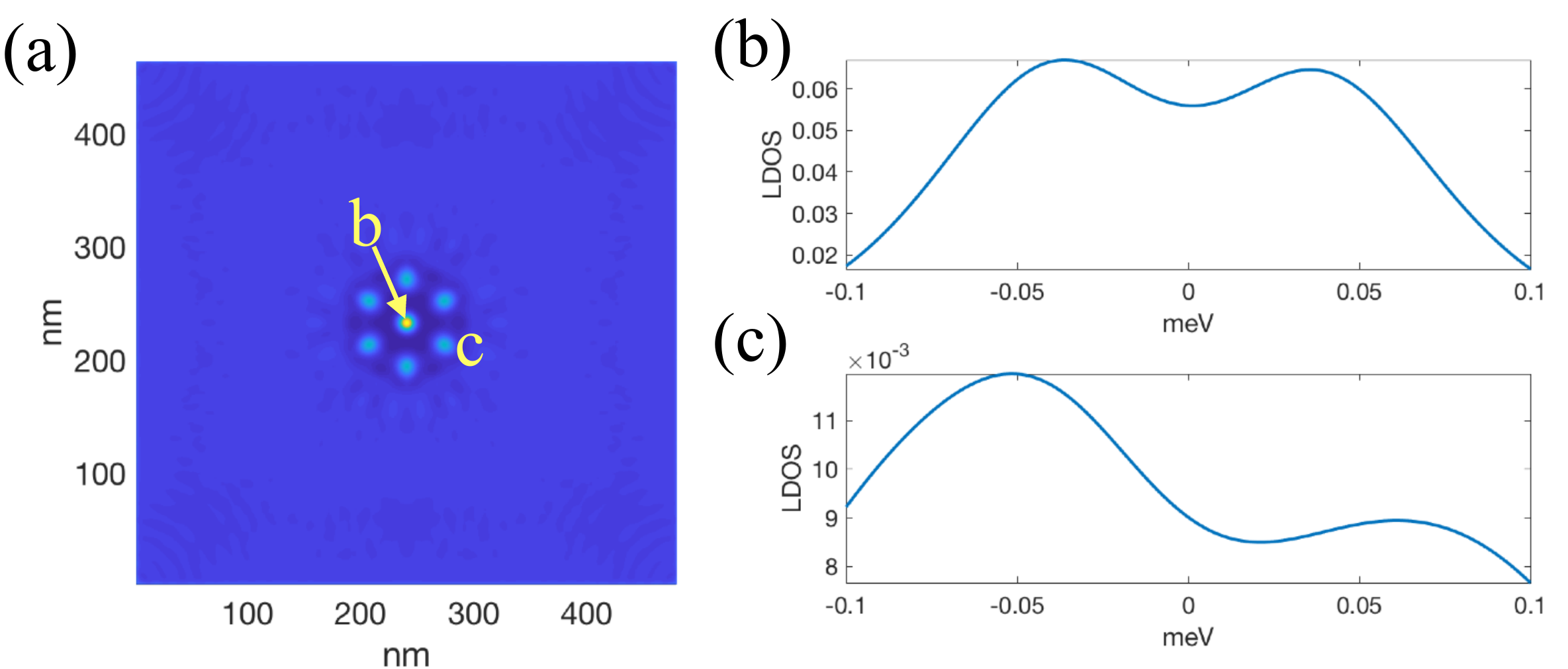}
\caption{ {\bf The TB simulation without magnet flux.} The vortices have perfect triangular lattice configuration. (a) shows LDOS with energy within $\pm0.12$meV and the localized low energy states appearing only near the center of the FOV. (b,c) show each vortex core possesses distinct LDOS so that the translational symmetry of the triangular lattice is completely broken.    }
\label{lpdinf} 
\end{center}
\end{figure}

	Choosing the size ($L_x,L_y$) of the $(001)$ surface is subtle. Ideally, the lengths of the surface edges should be much greater than the Majorana coherence length ($\xi=13.9$nm) and the London penetration depth ($\lambda=500$nm). That is, the surface size for the simulation has to be at least roughly greater than $1000\times 1000$ in the unit of the lattice constant ($a=2$nm). It will cost tremendous computer resources to solve the TB model in this size. Instead, we judiciously circumvent this scaling problem. We consider two types of surface sizes for the simulation --- real size and virtual size. The real size indicates the range including almost all of the MVM couplings; the couplings outside this range, which exponentially decay, can be neglected. In other words, the range of the summations in the BdG Hamiltonian (\ref{main BdG Hamiltonian}) is given by $|r^x|<L_x^r/2$ and $|r^y|<L_y^r/2$, where $(r^x,r^y)=\br$.  Hence, as shown in Fig.~\ref{simulation_size}(b) we choose the real size $L_x^r\times L_y^r=464\times 480$nm, which is much greater than the Majorana coherence length ($13.9$nm). On the other hand, the virtual size indicates the range including almost all of the phase effect from the vortices. That is, the coupling of the two nearby MVMs is affected by phases from far away vortices. When the distance between the coupled MVMs and the far-away vortices is much greater than the London penetration depth, the phase effect is screened out by Eq.~\ref{screening}. Therefore, the virtual size must be greater than the real size and the London penetration depth so that the phase effect for the Majorana coupling in the real size range is included in the virtual size range. As shown in Fig.~\ref{simulation_size}(b), we choose virtual size $L_x^v\times L_y^v=3712\times 3840$nm. In other words, the range of the summation in the order parameter (\ref{Delta}) and the vector potential (\ref{vector potential all}) are given by $|r^x_j|<L_x^v/2$ and $|r^y_j|<L_y^v/2$, where $(r^x_j,r^y_j)=\br_j$.

\begin{figure}[t!]
\begin{center}
\includegraphics[clip,width=0.99\columnwidth]{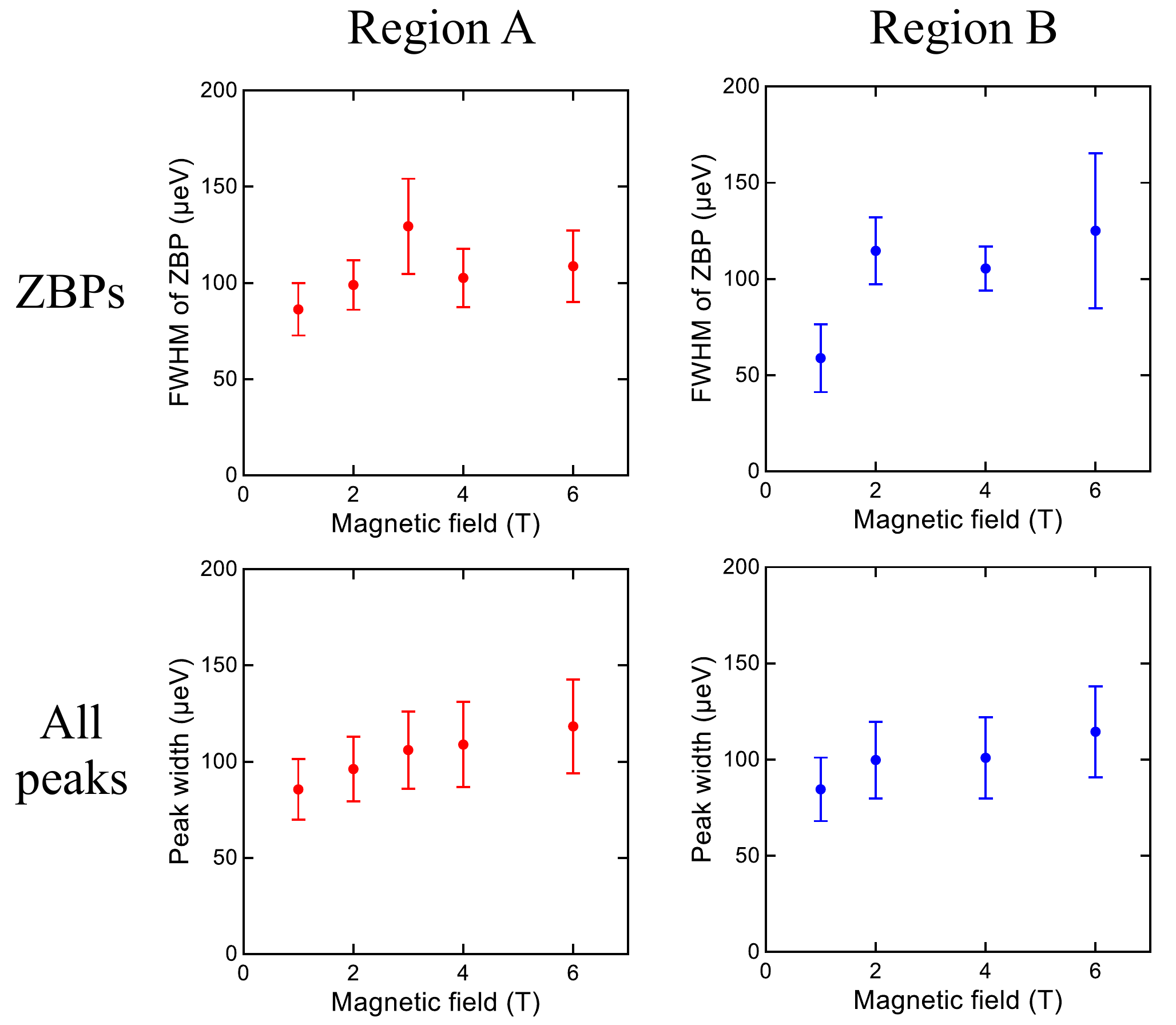}
\caption{ {\bf FWHM of the  tunneling peaks in the STM experiment~~\cite{2018arXiv181208995M}} Regions A and B represent two different experimental data sets. The averaged value of the FWHM in $B=1$T is around $80\mu$eV. 
 }
\label{supp_FWHM} 
\end{center}
\end{figure}

	Another time-consuming process for the simulation is to compute the integral $\frac{e}{\hbar}\int_{\br}^{\br+\bm{\delta}} \bm{A}(\br)\cdot d\bm{l}$ for each hopping term in the BdG Hamiltonian (\ref{main BdG Hamiltonian}) because without the vector potential the TB simulation is incorrect as shown in Fig.~\ref{lpdinf}. Since the distance $\delta$ of the two nearest neighbor atoms is small, we use this approximation 
\bee
\frac{e}{\hbar}\int_{\br}^{\br+\bm{\delta}} \bm{A}(\br)\cdot d\bm{l}\approx \sum_{j}\theta_j (1-\frac{|\bar{\br}-\br_j|}{\lambda}K_1(\frac{|\bar{r}-r_j|}{\lambda})),
\ee
where $\bar{\br}=\br+\delta/2$, $\theta_j$ is the angle between line $\br_j,\br$ and line $\br_j,\br+\bm{\delta}$.  

	By using the workstation with 256GB memory and Xeon Gold 6130 processor, we perform the Lanczos algorithm to find the energy states roughly within $\pm 0.25$meV energy range in the BdG Hamiltonian (\ref{main BdG Hamiltonian}). Consequently, we have the states with energy $E_j$ close to zero in the form of 
\bee
\ket{\phi^j_\br}=(\bm{u}^j_\br\  \bm{\nu}^j_\br),
\ee
where $\bm{u}^j_\br, \bm{\nu}^j_\br$ are 4-component vector. These states can reveal the hybridization of multiple MVMs. To connect the physical observable, by using these states we can compute the LDOS on the top surface of ${\rm FeTe}_{x}{\rm Se}_{1-x}$ 
\bee
N(\br,E)= \sum_{E_j<0} \Big ( |\bm{u}^j_\br|^2  \delta(\beta (E-E_j))+ |\bm{\nu}^j_\br|^2  \delta(\beta(E+E_j)) \Big), \label{LDOSeq}
\ee
where $\delta(\beta E)=1/\cosh^2 (\beta E)$. We choose $1/\beta=k_BT=20\mu$eV for the simulation leading to $76\mu$eV FWHM of the tunneling peaks. This value of the FWHM is close to the averaged value of the FWHM (80$\mu$eV) in the experiment~\cite{2018arXiv181208995M} in the low magnetic field as shown in Fig.~\ref{supp_FWHM}.




\section{Particle-hole symmetry broken by Majorana hybridization} \label{PH symmetry broken}

For an isolated MVM with zero energy, its LDOS exhibits perfect particle-hole symmetry. On the other hand, Fig.~\ref{triangular_lattice}(c) shows that when the intervortex distance of the Majorana triangular lattice becomes shorter, the LDOS of the vortex exhibits the broken particle-hole symmetry. The reason is that the Majorana hybridization and wavefunction overlapping break particle-hole symmetry. To capture the relation between the Majorana physics and particle-hole symmetry, we start with two isolated MVMs with zero energy located at $\br_1$ and $\br_2$
\begin{align}
\ket{\phi_1(\br)}&=\gamma_1(\br) \ket{\rm{BCS}}, \\
\ket{\phi_2(\br)}&=\gamma_2(\br) \ket{\rm{BCS}}, 
\end{align}
where $\gamma_1(\br)=\bmu_1(\br)c^\dagger_\br +\bmu_1^*(\br) c_\br$, $\gamma_2(\br)=\bmu_2(\br)c^\dagger_\br +\bmu_2^*(\br) c_\br$, and $\ket{\rm{BCS}}$ is the BCS wavefunction. Here we use $\bnu_i^*=\bmu_i$ since a MVM with zero energy is its own antiparticle. By Eq.~\ref{LDOSeq}, each MVM has LDOS in the form of 
\bee
N_i(\br,E)=2 |\bmu_i(\br)|^2 \delta (\beta E).
\ee
By definition, the peak is located at zero energy and the LDOS obeys $N_i(\br,E)=N_i(\br,-E)$ indicating preserving particle-hole symmetry. As these two MVMs couple weakly, the Majorana wavefunctions are kept in the original forms, and the hybridization Hamiltonian can be generally written in the second quantization form 
\bee
H_h=it \gamma_1 \gamma_2.
\ee
By defining the BCS wavefunction satisfying $(\gamma_1+i\gamma_2)\ket{\rm{BCS}}=0$, we have an eigenstate $(\gamma_1-i\gamma_2)\ket{\rm{BCS}}$ with energy $E=-t$. The explicit form of the wavefunction is written as
\bee
 (\gamma_1-i\gamma_2){\ket{\rm{BCS}}}=\big( \bar{\bmu} (\br)c^\dagger_\br +\bar{\bnu}(\br) c_\br \big)\ket{\rm{BCS}},
 \ee
 where $\bar{\bmu}(\br)=\bmu_1(\br)+i\bmu_2(\br)$ and $\bar{\bnu}(\br)=\bmu_1^*(\br)+i\bmu_2^*(\br)$. The LDOS of the Majorana hybridized state is given by 
 \begin{align}
\bar{N}(\br,E)=&|\bar{\bmu}(\br)|^2\delta(\beta(E+t))+|\bar{\bnu}(\br)|^2\delta(\beta(E-t)).
\end{align}

\begin{figure}[t!]
\begin{center}
\includegraphics[clip,width=0.7\columnwidth]{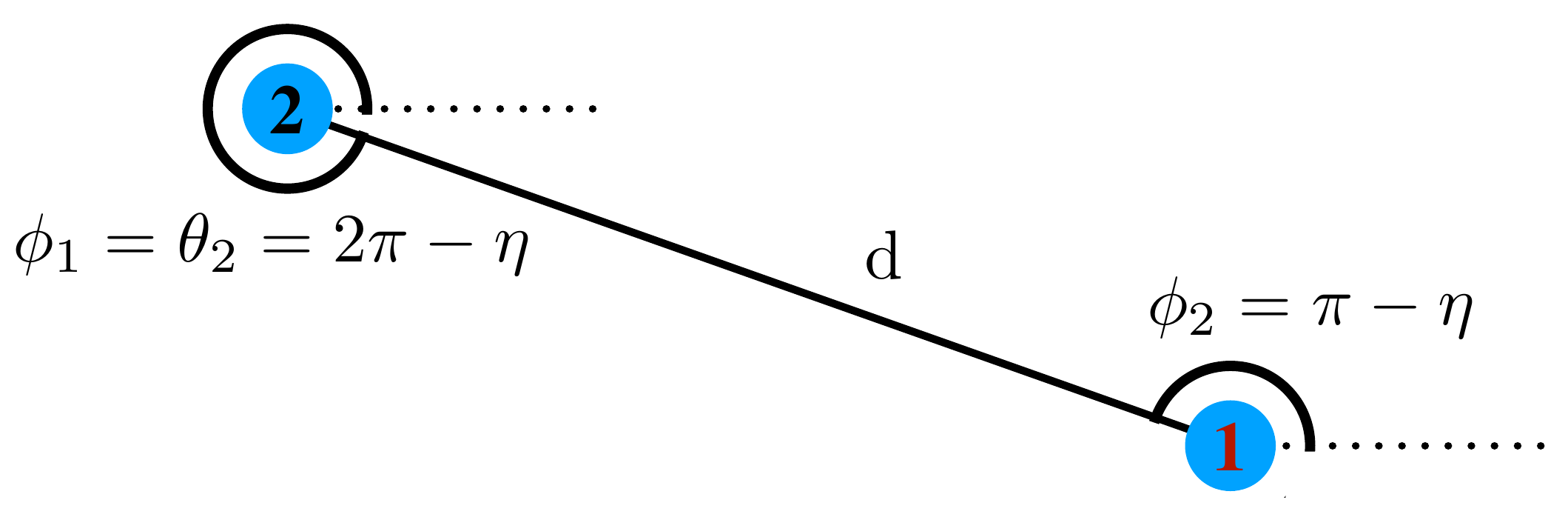}
\caption{ {\bf The angles and phases for a point near vortex $\#1$} It is unnecessary to know the value of $\theta_1$, since the LDOS of the hybridized MVMs near $\br_1$ is independent of the value of $\theta_1$. 
 }
\label{PH_symmetry_breaking} 
\end{center}
\end{figure}

This is the general form of the LDOS. Now we consider the specific case that the hybridization of the two MVMs residing at two vortex cores ($\br=\br_1,\ \br_2$) respectively. 
In the Hamiltonian (\ref{cont model}), the two phases $(\theta_i,\phi_i)$ of the order parameter stem from the phase winding of vortex $\# i$ and other vortices respectively so that $\Delta = \Delta_0 e^{i(\theta_i+\phi_i)}$. By definition, we have $\theta_i+\phi_i=\theta_j+\phi_j$. Here we assume $\phi_i(r)$ is a constant for MVM $\# i$ since for vortex $\# i$ $\theta_i$ is used in the polar coordinate to solve the eigenvalue problem. With the additional phase $\phi_i$, the wavefunction (\ref{cont state}) of the MVM $\# i$ has a modification  
\bee
u_i(\br)=e^{-|\br-\br_i|/\xi+i \phi_i/2} \big( J_0 (k_F|\br-\br_i|), e^{i\theta_i} J_1 (k_F|\br-\br_i|) ).
\ee
Consider the LDOS at one of the vortex core, say $\br= \br_1$. We simplify the problem by neglecting the magnetic flux effect ($\bm{A}(\br)$); the distribution of the two vortices gives the specific values of the three phases $\phi_1=- \eta,\ \theta_2=-\eta,\ \phi_2=\pi-\eta$ as illustrated in Fig.~\ref{PH_symmetry_breaking}, where $\eta$ is an arbitrary angle. We have the explicit form of the hybridized wavefunction
\begin{align}
\bar{u}(\br_1)=& e^{-i\eta/2}\big(1-e^{-d/\xi}J_0(k_Fd) , -e^{-d/\xi} J_1(k_Fd)e^{-i\eta}\big), \nonumber \\
\bar{\nu}(\br_1)=&e^{i\eta/2}  \big( 1+ e^{-d/\xi}J_0(k_Fd), e^{-d/\xi} J_1(k_Fd))e^{i\eta} \big).
\end{align}
where $d\equiv |\br_1-\br_2|$ is the distance between the two MVMs. Here we use $J_0(0)=1,\ J_1(0)=0$.  The LDOS at $\br=\br_1$ can be separated to two parts 
 \begin{align}
\bar{N}(\br_1,E)=&N_{\rm{e}}(E)+N_{\rm{o}}(E),
\end{align}
where 
\begin{align}
N_{\rm{e}}(E) =& \Big [  1 + e^{-2d/\xi } \big( J_0^2(k_Fd) +J_1^2(k_Fd) \big)\Big]  \nonumber \\
&\times \big[\delta(\beta(E+t))+\delta(\beta(E-t)) \big],  \\
N_{\rm{o}} (E)=&2 e^{-d/\xi } J_0(k_Fd) \big[-\delta(\beta(E+t))+\delta(\beta(E-t)) \big].
\end{align}
While the even function $N_{\rm{e}}(E)$ of $E$ preserves particle-hole symmetry, the odd function $N_{\rm{o}}(E)$ of $E$ breaks particle-hole symmetry. 
Due to the exponential decay $e^{-d/\xi }$, the distance between the two MVMs controls the strength of the particle-hole symmetry breaking. This idea can be extended to the relation between the intervortex distance and the particle-hole symmetry breaking in the Majorana lattice system. Hence, the short intervortex distance results in broken particle-hole symmetry in the vortex core as shown in Fig.~\ref{triangular_lattice}(c).



\begin{widetext}
\newpage 

\section{Simulation for a moving vortex}
	
	Fig.~\ref{vortex_move} in the main text shows the LDOS changes of the vortex movement only in the high magnetic field ($\approx 4$T). Here we further show the vortex movement in two lower magnetic fields ($1$T and $2$T) as shown in Fig.~\ref{distribution46_5nm},\ref{distribution33nm}. The relation between the Majorana hybridization and the moving   vortex is still complicated. For example, as shown in Fig.~\ref{distribution33nm} the vortex core $\delta$, which originally does not possess a ZBP, has a ZBP after vortex $\alpha$ moves away from vortex $\delta$. However, the overall behavior of the LDOS is similar to the high magnetic field. That is, the LDOS of the vortices near the moving vortex has significant changes. For vortices away from the moving vortex, the LDOS exhibits no change (Fig.~\ref{distribution46_5nm}$\delta$) or slight change (Fig.~\ref{distribution33nm}$\epsilon$).

\begin{figure}[b!]
\begin{center}
\includegraphics[clip,width=0.54\columnwidth]{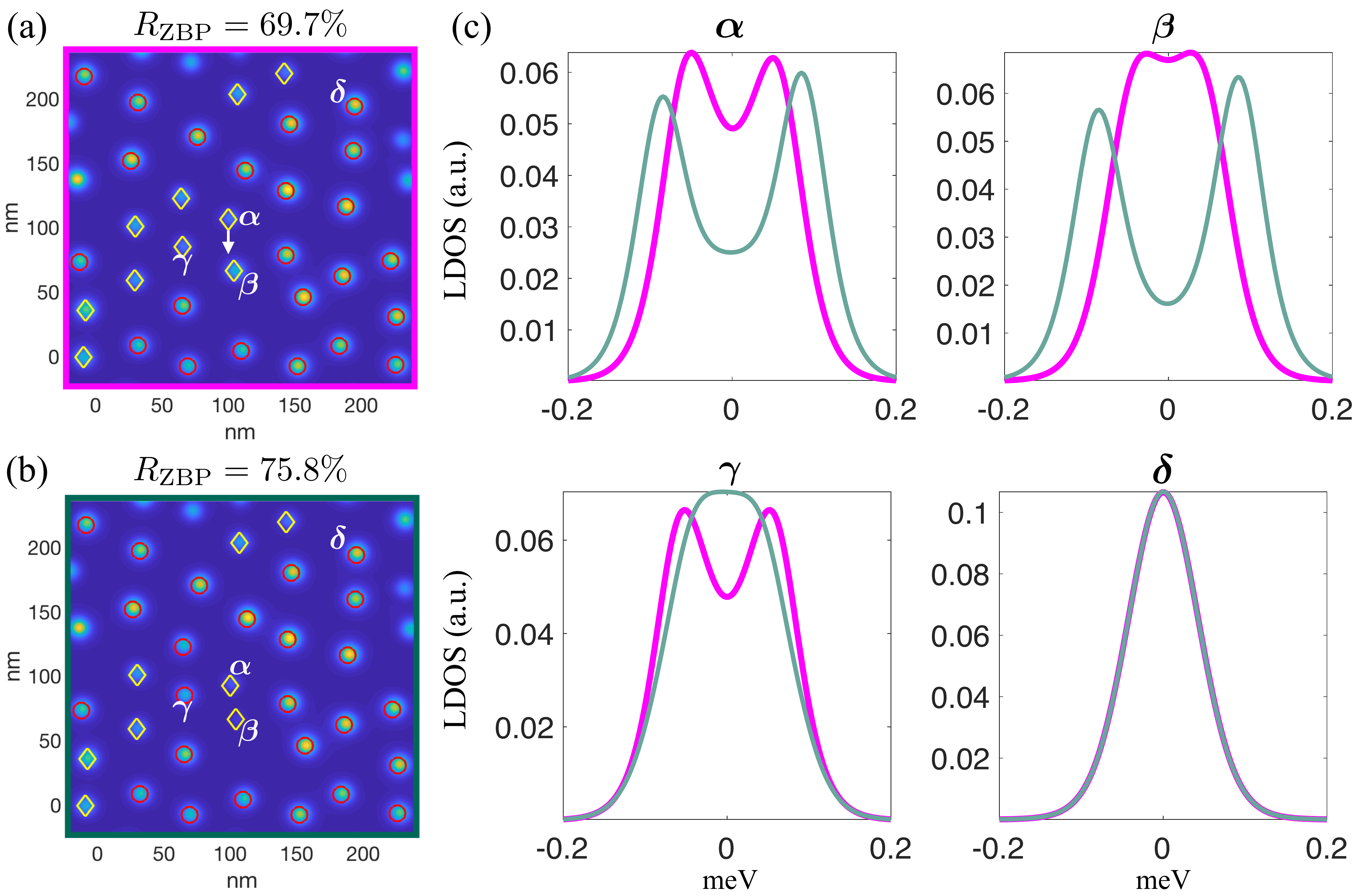}
\caption{ {\bf The LDOS changes with $\bar{R}_v=46.5$nm corresponding to $B=1$T before and after the central vortex $\alpha$ moves $14$nm down } (a,b) show LDOS with energy within $\pm 37\mu$eV. The vortex arrangement in the low magnetic field follows distorted lattice distribution, which in Fourier space is similar with Fig.~\ref{ZBP_distribution}(b). The red circles and yellow diamonds indicate the vortex cores with/without ZBPs respectively.  (c) shows the LDOS for four selected vortex cores before (pink) and after (green) the vortex movement. 
 }
\label{distribution46_5nm} 
\end{center}
\end{figure}

\begin{figure*}[b!]
\begin{center}
\includegraphics[clip,width=.6\columnwidth]{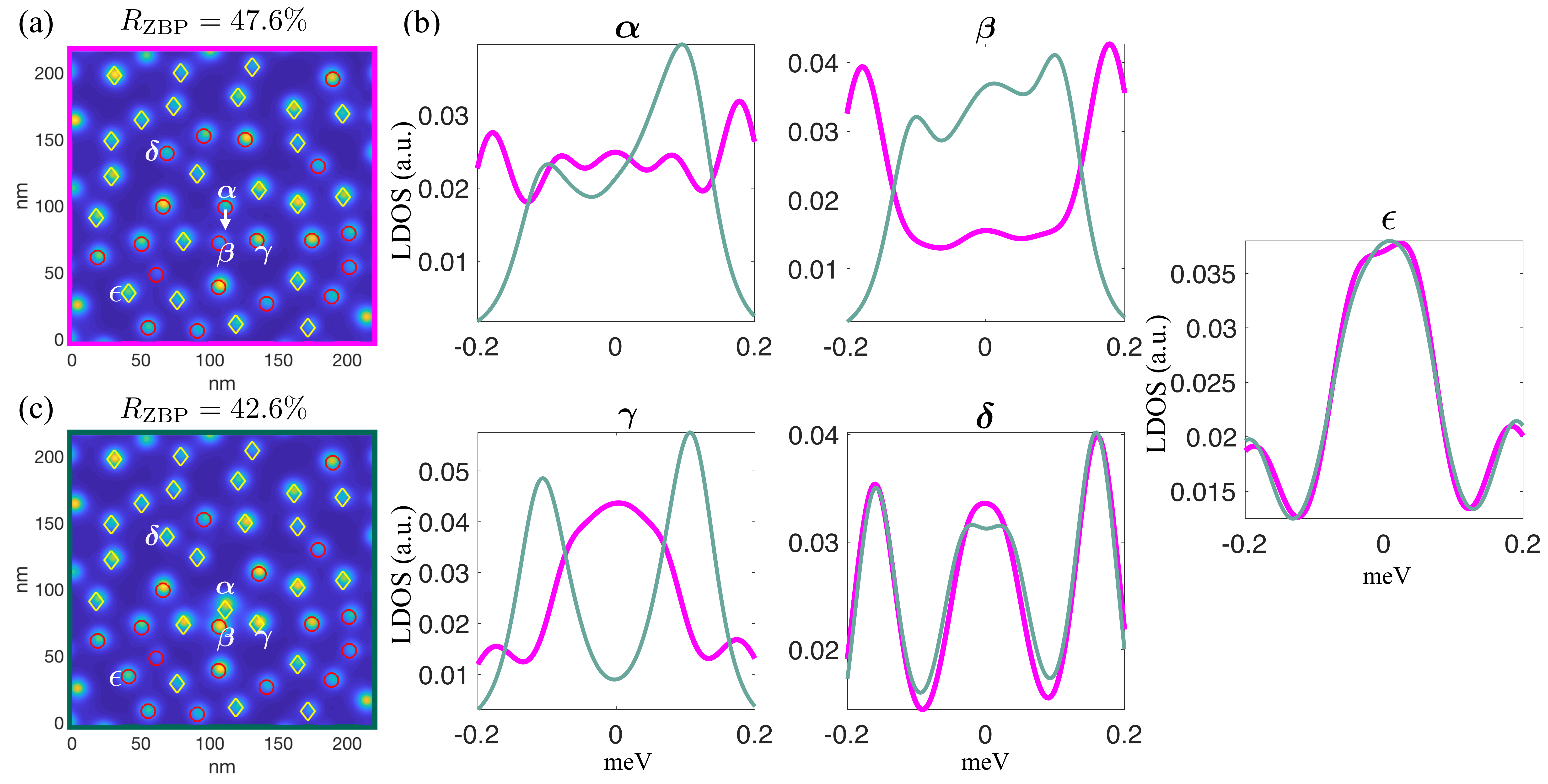}
\caption{ {\bf The LDOS changes with $\bar{R}_v=33$nm corresponding to $B=2$T before and after the central vortex $\alpha$ moves $14$nm down } (a,b) show LDOS with energy within $\pm 0.14$meV. The vortex arrangement in the low magnetic field follows glass distribution, which in Fourier space is similar with Fig.~\ref{ZBP_distribution}(f). The red circles and yellow diamonds indicate the vortex cores with/without ZBPs respectively.  (c) shows the LDOS for five selected vortex cores before (pink) and after (green) the vortex movement. 
 }
\label{distribution33nm} 
\end{center}
\end{figure*}

\newpage 

\section{STM measurement of creeping vortex and vortex relocation}
	
	Fig.~\ref{creeping_vortex} shows the LDOS of the creeping vortex has a dramatic change after this vortex moved $2$nm. The range of the FOV is $15\times 15$nm, which is comparable to the Majorana coherence length ($\xi=13.9$nm). Therefore, it is possible that the surrounding vortices, which were not seen by the STM, also moved to affect the LDOS of the creeping vortex. Fig.~\ref{experiment_4T_vortex_redistribution} shows the LDOS of the vortex core was not affected by the vortex relocation, since the shortest distance between the probed vortex and the visible relocated vortices is roughly $50$nm ($> \xi$). 

\begin{figure}[b]
\begin{center}
\includegraphics[clip,width=0.70\columnwidth]{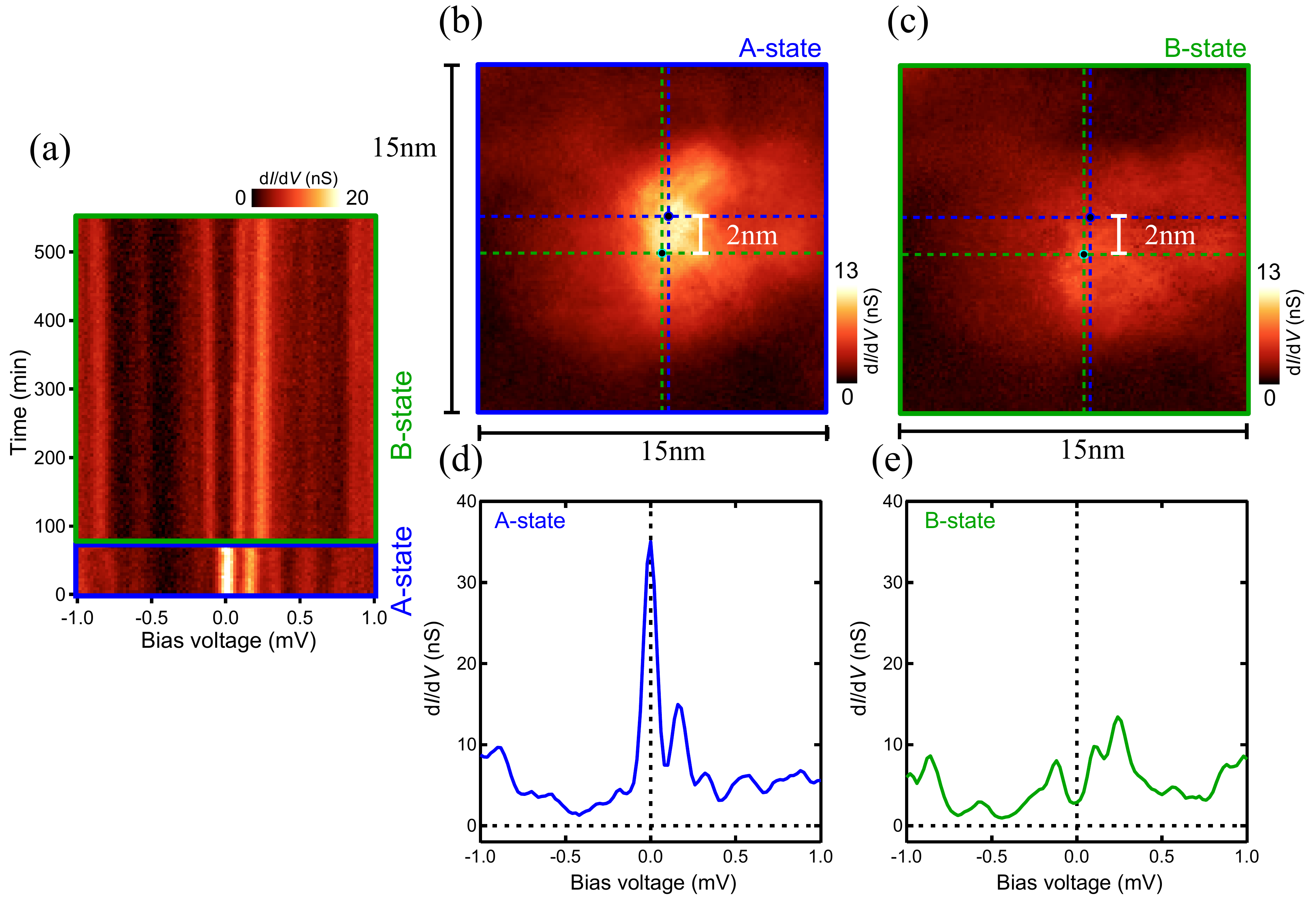}
\caption{ {\bf Time dependence measurements on a creeping vortex in $B=4$T with averaged intervortex distance $\bar{R}_v\approx 23$nm}   (a) shows the tunneling spectrum of the same vortex core as the time varies. (b,c) zero-energy conductance maps in the FOV before and after the vortex movement. The vortex moved $2$nm from the blue crossing to the green crossing at 70mins. (d,e) show the tunneling spectra at the location of the vortex core in the A-state, before and after the creep. (The location of the STM tip is fixed at the position of the vortex core in the A-state.)  
 }
\label{creeping_vortex} 
\end{center}
\end{figure}

\begin{figure}[b!]
\begin{center}
\includegraphics[clip,width=.7\columnwidth]{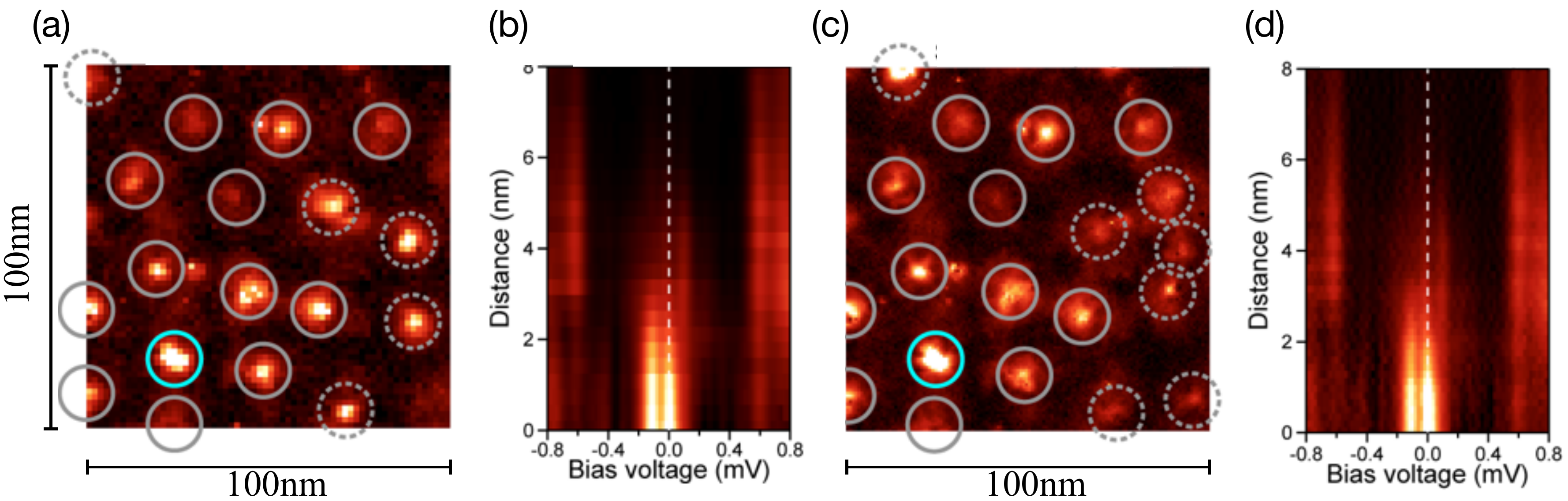}
\caption{ {\bf The tunneling spectra before and after the vortex relocation in the same FOV} (a,b) were taken after the first ramping in $B=4$T with averaged intervortex distance $\bar{R}_v\approx 23$nm, while (c,d) were taken after the second ramping. (a,c) zero-energy conductance maps before and after the vortex relocation. While the solid circles indicate the pinned vortices, the dashed circles indicate the relocated vortices. (b,d) line profiles of the tunneling spectra in the left lateral direction from the probed vortex core in the blue circle before and after the vortex relocation.  Since these two spectra are almost identical, the LDOS of this probed vortex was not significantly affected by the vortex relocation. The reason is that the shortest distance between the probed vortex and the visible relocated vortices ($\sim 50$nm) is much greater than the Majorana coherence length ($13.9$nm).    
 }
\label{experiment_4T_vortex_redistribution} 
\end{center}
\end{figure}

\newpage

\section{Simulation for LDOS line profiles of ZBP and non-ZBP vortices}

	Fig.~\ref{spin_LDOS} shows (spin) LDOS line profiles of the two selected vortices in Fig.~\ref{ZBP_distribution}. As shown in panel (a), the spatial length ($5$nm) of the ZBP is consistent with the STM experiment observation of ${\rm FeTe}_{x}{\rm Se}_{1-x}$~\cite{Wang2018Evidence,2018arXiv181208995M}. In addition, this observable length is identical to the spatial length of the splitting energy peaks as shown in panel (b). Panel (c,d) show that the spin up line profiles with and without ZBPs are almost the same with the line profile including both spins in panel (a,b).  The spin down line profiles in panel (e,f) show that the spatial peaks appear away from the vortex cores regardless of the splitting of the energy peaks. The heights of the spin up peaks are roughly 6 times smaller than the ones of the spin up.   


\begin{figure}[b]
\begin{center}
\includegraphics[clip,width=0.69\columnwidth]{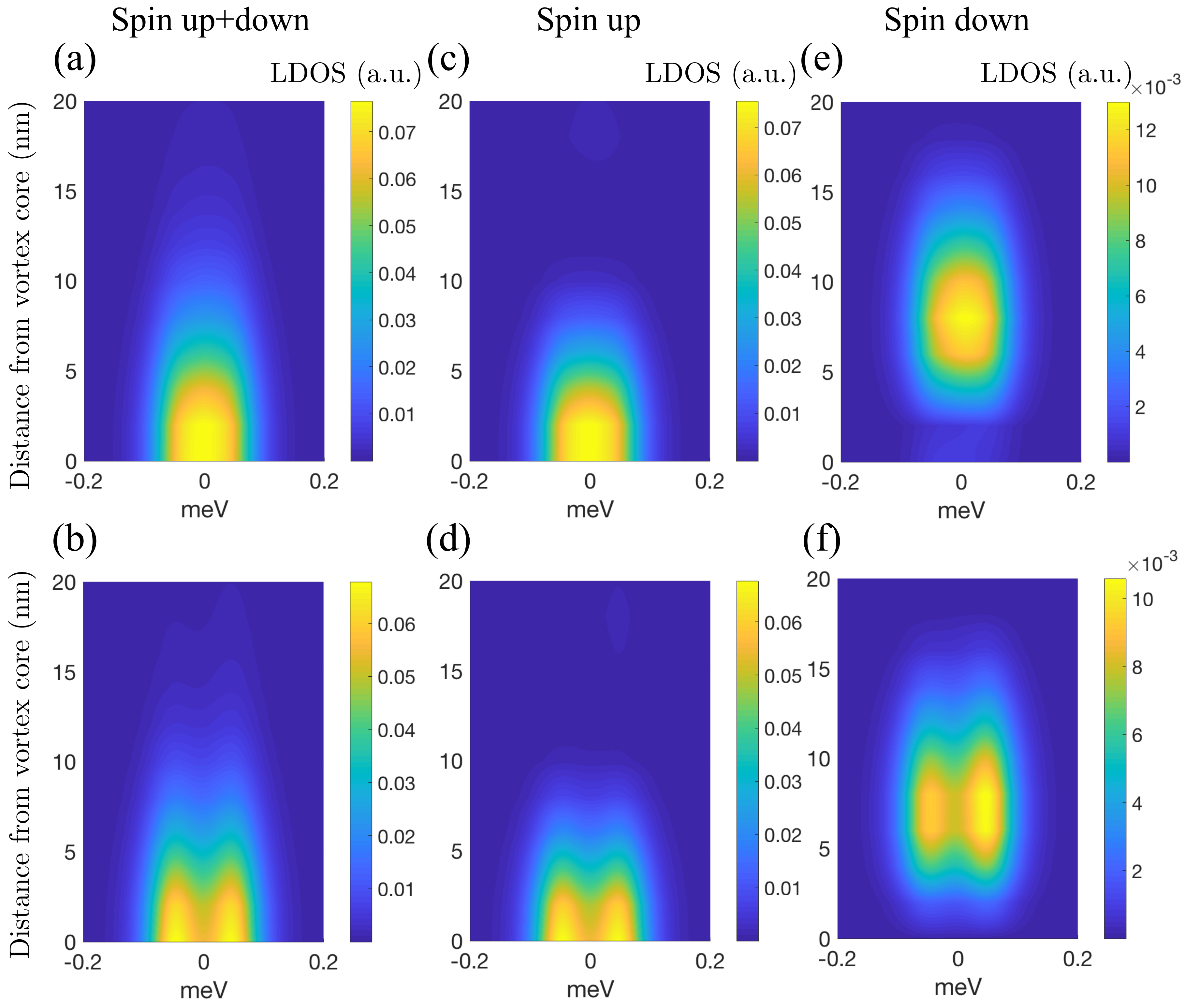}
\caption{ {\bf (Spin) LDOS line profiles of the two selected vortices (c,d) in Fig.~\ref{ZBP_distribution}(a) with averaged intervortex distance $\bar{R}_v\approx 46.5$nm } All of the line profiles start from the vortex cores in the up lateral direction in Fig.~\ref{ZBP_distribution}(a). (a,c,e) show (spin) LDOS of vortex c possessing a ZBP (b,d,f) show (spin) LDOS of vortex d possessing two energy splitting peaks away from zero energy.  
 }
\label{spin_LDOS} 
\end{center}
\end{figure}

\newpage

\section{Inhomogeneous superconducting gaps and chemical potentials} \label{inhomogeneous}

It has been observed that the (001) surface of ${\rm FeTe}_{x}{\rm Se}_{1-x}$ has inhomogeneous superconducting gap distribution (see supplementary materials in~\cite{2018arXiv181208995M}). We investigate whether this inhomogeneity affects the Majorana hybridization by examining LDOS of vortex cores in triangular vortex lattice. We use the spatial distribution of the superconducting gap based on the supplementary materials in~\cite{2018arXiv181208995M} for the simulation. Fig.~\ref{disorder_gap} shows in the presence of the inhomogeneous superconducting gaps each vortex in the same magnetic field has almost identical LDOS. Hence, the inhomogeneous gaps do not seriously affect the Majorana hybridization on the surface of ${\rm FeTe}_{x}{\rm Se}_{1-x}$. On the other hand, the inhomogeneous chemical potentials significantly affect the Majorana hybridization. As shown in Fig.~\ref{disorder_cp} this inhomogeneity leads to different LDOS for each vortex core in the low and high magnetic fields. 

\begin{figure*}[b]
\begin{center}
\includegraphics[clip,width=0.80\columnwidth]{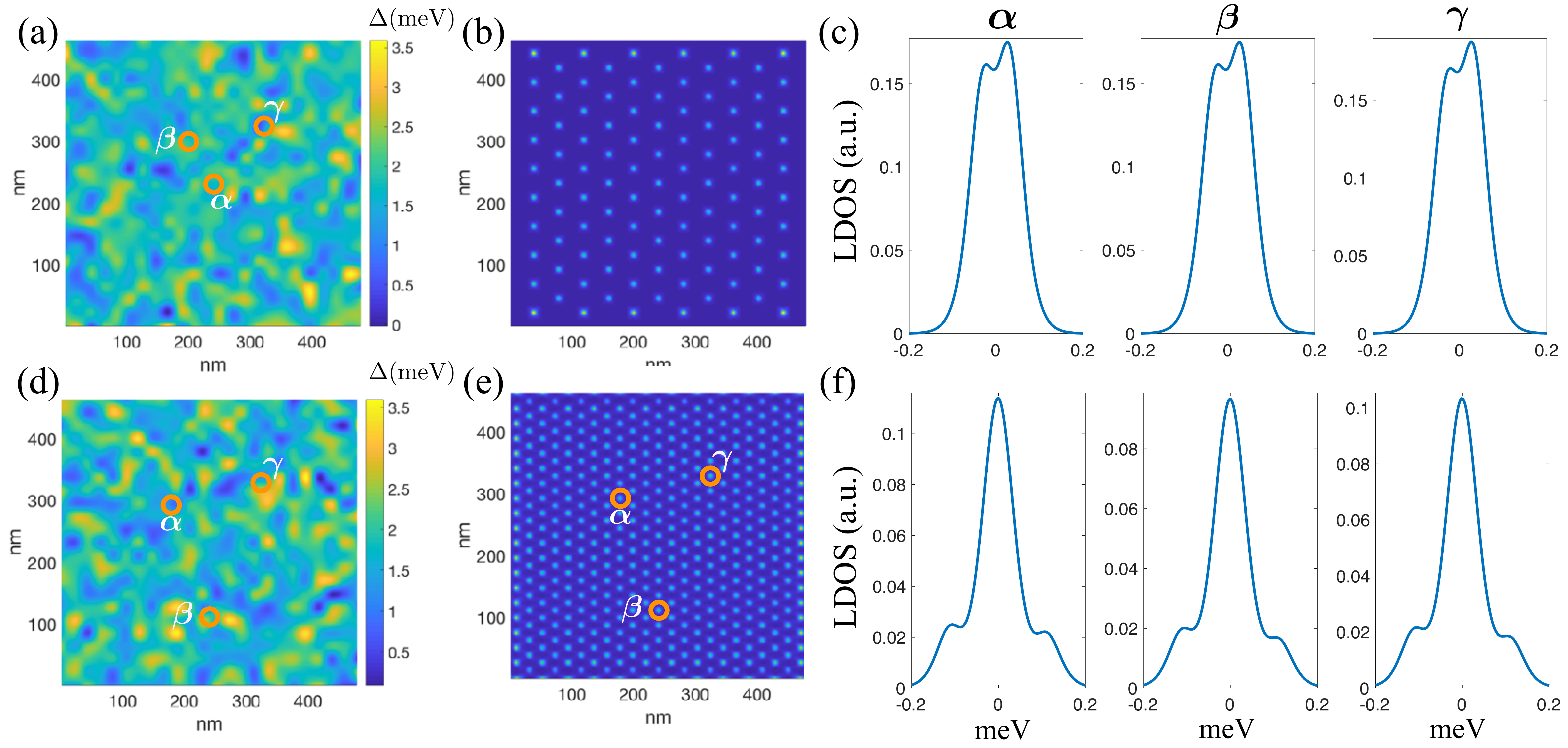}
\caption{ {\bf Inhomogeneous superconducting gaps for triangular vortex lattice with intervortex distances (a-c) $46.5$nm and (d-f) $24$nm} (a,d) show spatial distributions of the superconducting gaps, which are similar to the STM observation (supplementary information~\cite{2018arXiv181208995M}). (b,e) show LDOS with energy within $\pm 33\mu$eV and $\pm 0.12$meV respectively and triangular vortex distributions in the FOVs. (c,f) show the LDOS of the selected vortex cores. Although there are slight changes of the energy peak heights, the LDOS for all of the vortices is almost identical.  
 }
\label{disorder_gap} 
\end{center}
\end{figure*}

\begin{figure*}[b!]
\begin{center}
\includegraphics[clip,width=0.80\columnwidth]{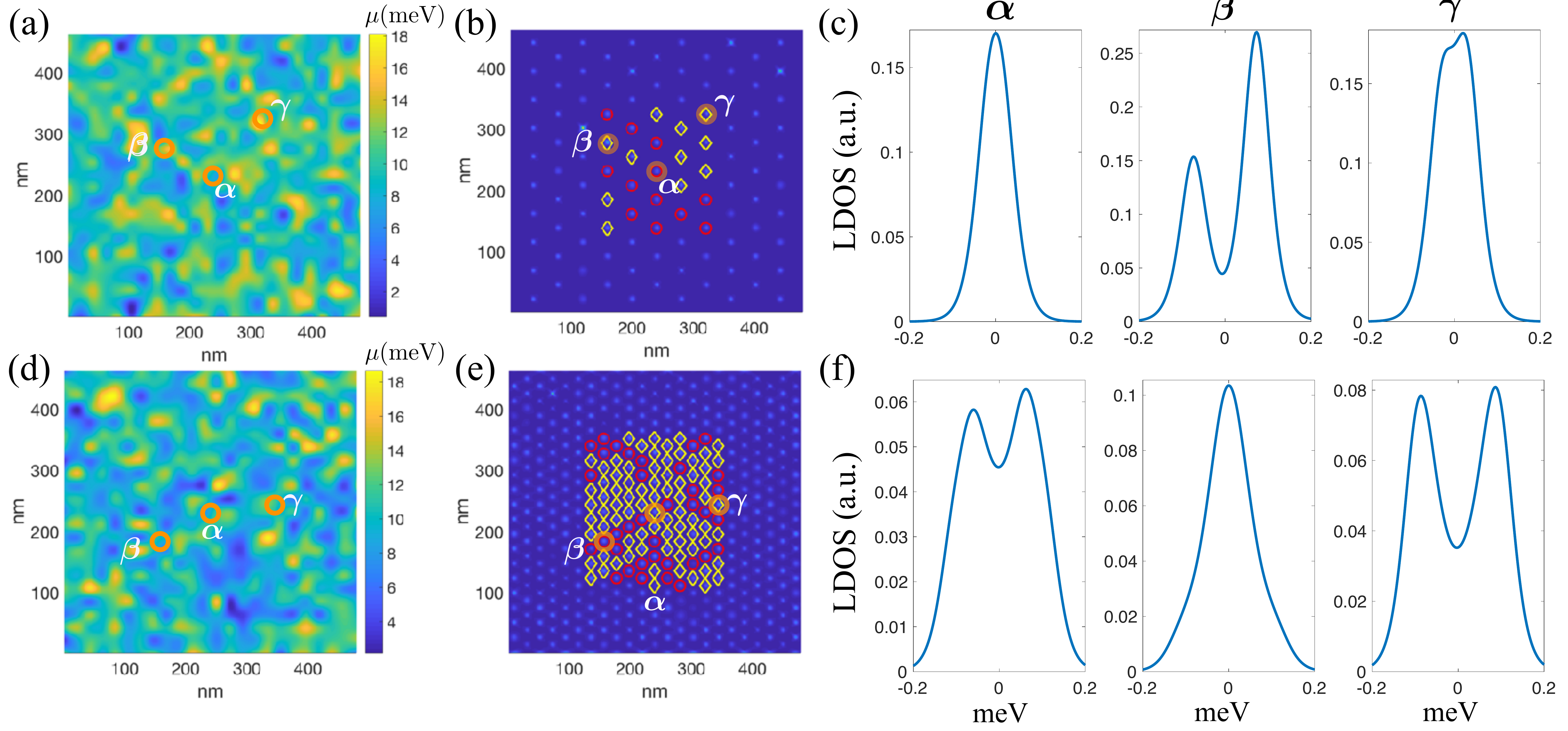}
\caption{ {\bf Inhomogeneous chemical potentials for triangular vortex lattice with intervortex distances (a-c) $46.5$nm and (d-f) $24$nm} (a,d) show spatial distributions of the chemical potentials with wide range changes. (b,e) show LDOS with energy within $\pm 0.36meV$ and $\pm 0.13meV$ respectively and triangular vortex distributions in the FOVs. The red circles and yellow diamonds indicate the vortex cores with/without ZBPs respectively.  (c,f) show the LDOS of the selected vortex cores.  Each of the selected vortices has distinct LDOS. 
 }
\label{disorder_cp} 
\end{center}
\end{figure*}


\newpage

\section{ZBP heights} \label{ZBP heights}

	As the magnetic field increases, the intervortex distance decreases so that the Majorana hybridization enhances. It is not surprising that the averaged ZBP height decreases with the increasing magnetic field. As shown in Fig.~\ref{ZBP_height_Exp}(a,b) the averaged ZBP heights of the LDOS in the vortex cores from the simulation exhibit this decreasing behavior. The experimental data only in region A have this similar behavior of the tunneling conductance as shown in Fig.~\ref{ZBP_height_Exp}(c). However, in Fig.~\ref{ZBP_height_Exp}(d) the data in region B do not exhibit deceasing ZBP heights. The reason of the inconsistency might be that in different magnetic fields the tunnelings of the STM were not in the same conditions. It is important to further experimentally check the decreasing averaged ZBP heights in ${\rm FeTe}_{x}{\rm Se}_{1-x}$. 

\begin{figure}[b!]
\begin{center}
\includegraphics[clip,width=.58\columnwidth]{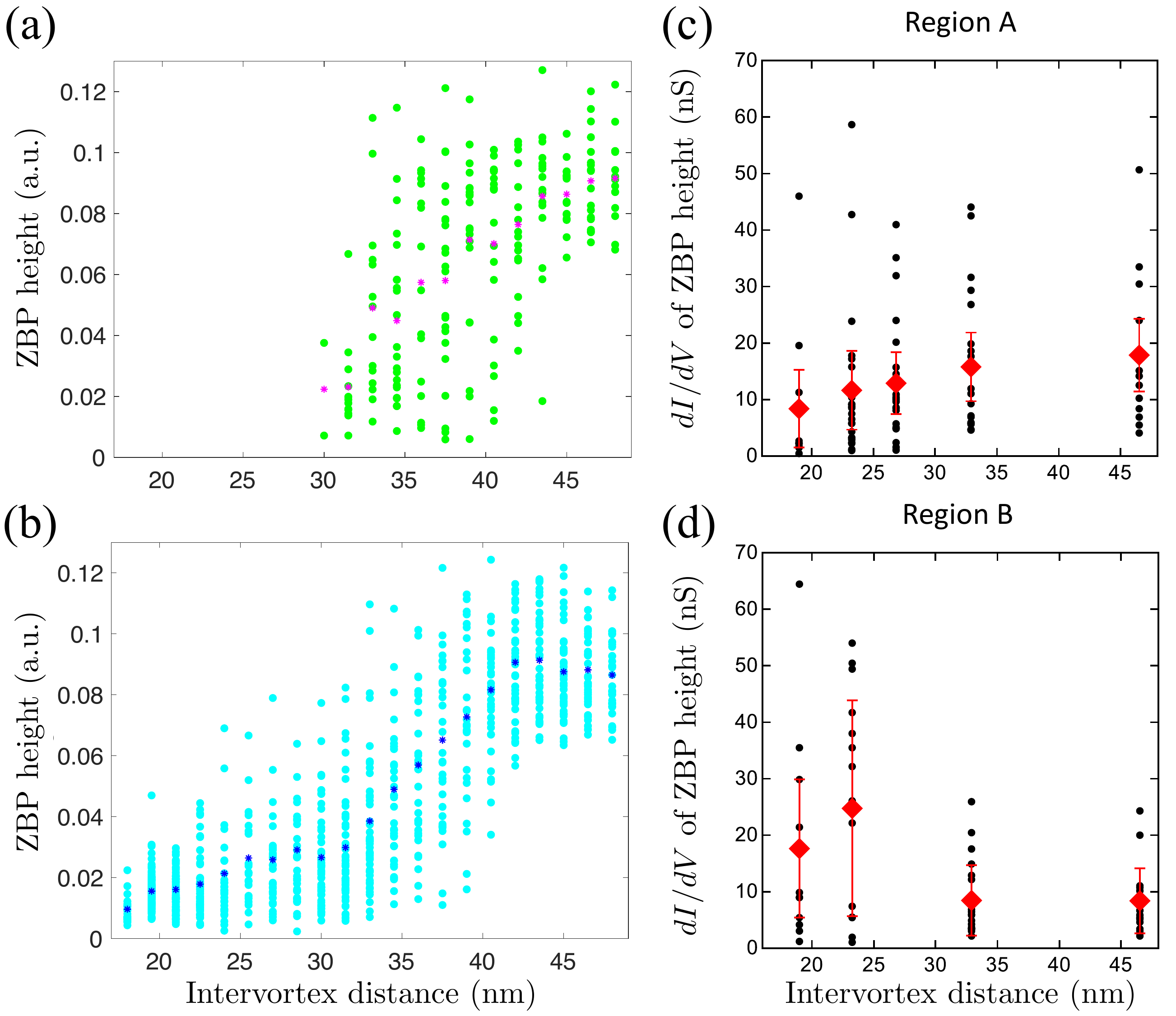}
\caption{ {\bf ZBP heights from the simulation (a,b) and the STM experiment (c,d) in different magnetic fields} (a) Green(magenta) dots indicate (averaged) ZBP heights in the vortex cores having distorted triangular lattice distribution on the surface of the iron-based superconductor. For the distorted vortex lattice, the ZBP rate is almost zero as the intervortex distance is less than $30$nm. (b) Cyan(blue) dots indicate (averaged) ZBP heights in the vortex cores having glass distribution. (c,b) show ZBP heights (black dots) and averaged heights (red diamonds) in the two sets of the experiment data~\cite{2018arXiv181208995M} (region A and B).  
 }
\label{ZBP_height_Exp} 
\end{center}
\end{figure}

\end{widetext}




\end{document}